\documentclass[12pt]{article}

\usepackage{epsfig}

\usepackage{amsmath}
\usepackage{amssymb}
\usepackage{euscript}

\newcommand{\Pcal}{{\cal P}}

\newcommand{\Peu}{\EuScript{P}}

\newcommand{\Deu}{\EuScript{D}}

\newcommand{\bu}{\bullet}
\newcommand{\veps}{\varepsilon}

\newcommand{\boldtau}{\mbox{\boldmath$\tau$}}

\begin{document}

\begin{titlepage}


\begin{flushright}
\bf IFJPAN-V-04-06\\
\bf CERN-PH-TH/2005-065
\end{flushright}

\vspace{1mm}
\begin{center}
  {\Large\bf%
    Non-Markovian Monte Carlo Algorithm for
    the Constrained Markovian Evolution in QCD$^{\star}$
}
\end{center}
\vspace{1mm}

\begin{center}
{\bf S. Jadach}
{\em and}
{\bf M. Skrzypek} \\

\vspace{1mm}
{\em Institute of Nuclear Physics, Polish Academy of Sciences,\\
  ul. Radzikowskiego 152, 31-342 Cracow, Poland}\\
and\\
{\em CERN Department of Physics, Theory Division\\
CH-1211 Geneva 23, Switzerland}
\end{center}

\vspace{5mm}
\begin{abstract}
We revisit the challenging problem of finding an efficient
Monte Carlo (MC) algorithm solving
the constrained evolution equations for the initial-state QCD radiation.
The type of the parton (quark, gluon) and 
the energy fraction $x$ of the parton exiting emission chain
(entering hard process) are predefined, i.e. constrained throughout the evolution.
Such a constraint is mandatory for any realistic MC for the initial state
QCD parton shower.
We add one important condition: the MC algorithm
must not require the {\em a priori} knowledge of the
full numerical exact solutions of the evolution equations,
as is the case in the popular ``Markovian MC for backward evolution''.
Our aim is to find at least one
solution of this problem that would function in practice.
Finding such a solution seems to be definitely within the reach of
the currently available computer CPUs
and the sophistication of the modern MC techniques.
We describe in this work the first example of 
an efficient solution of this kind.
Its numerical implementation is still restricted to the pure gluon-strahlung.
As expected, it is not in the class of the so-called Markovian MCs.
For this reason we refer to it as belonging to a class of
{\em non-Markovian} MCs.
We show that numerical results of our new MC algorithm
agree very well (to $0.2\%$) with the results of the other MC program
of our own (unconstrained Markovian) 
and another non-MC program {\tt QCDnum16}.
This provides a proof of the existence of the new class of MC techniques, 
to be exploited in the precision perturbative QCD calculations
for the Large Hadron Collider.
\end{abstract}

\vspace{3mm}
\begin{center}
\em To be submitted to Acta Physica Polonica
\end{center}

\vspace{4mm}
\begin{flushleft}
{\bf IFJPAN-V-04-06\\
\bf CERN-PH-TH/2005-065}
\end{flushleft}

\vspace{5mm}
\footnoterule
\noindent
{\footnotesize
$^{\star}$Supported in part by the EU grant MTKD-CT-2004-510126,
  in partnership with the CERN Physics Department.
}

\end{titlepage}

\section{Introduction}
The unprecedented experimental precision
of the forthcoming experiments at the Large Hadron Collider (LHC),
in terms of apparatus resolution
and event statistics, will have to be matched by a far better
precision of the theoretical calculations in the strong interaction sector
than available at present.
The well established theory of strong interactions,
Quantum Chromodynamics (QCD), is in principle able to provide
very precise predictions for the high energy scale 
(mass, transverse momentum, momentum transfer) processes.
The perturbative predictions of QCD are obtained within one of two
very different calculational frameworks:
the so-called {\em matrix element} (ME)
calculations and models of the {\em parton shower}
(PS) type implemented in the Monte Carlo (MC) event generators.
For a more detailed review of these methods,
see for example ref.~\cite{Catani:2000jh}.
In the ME calculations the basic ingredients are real- and virtual-emission
matrix elements evaluated in the fixed-order perturbative QCD,
for the hard process at the high energy scale,
embedded in the standard Lorenz-invariant phase space (LIPS).
The fixed-order ME is combined with the parton distributions (PDFs)
describing lower energy multiple emissions in an inclusive manner
(integrated over the transverse momenta).
On the other hand, the PS framework
offers a fully exclusive picture, down to hadronization energy scale,
that is the true MC events with explicit 4-momenta,
for all multiple soft and collinear emissions associated with the hard
process -- the same emissions
as are encapsulated in the PDFs of the ME approach.
However, the classic PS implements the hard process only 
at the Born (tree) level.

The above two complementary approaches have their strong and weak points
of their own.
Without entering into details, we may safely say
that it is absolutely mandatory to combine the
virtues of the two approaches if one hopes to ever achieve a significant
improvement of the precision of the QCD predictions,
for a wide class of observables (not only total rates);
see conclusions of ref.~\cite{Catani:2000jh}.

There were numerous attempts to combine ME calculations with the
parton shower approach beyond the leading order, the most elaborate
being the recent one of Frixione and Webber~\cite{Frixione:2002ik}.
However, none of them are fully satisfactory and there are more
proposals in this direction; see for instance ref.~\cite{Nason:2004rx}.
There seems to be a growing consensus that part of the problem 
is in the fundamental formulation of the PS models implemented in the PS MC.
All these models are of the Markovian%
\footnote{Since the adjective ``Markovian'' is (ab)used for a wide range of the
  phenomena, let us state that we understand by the Markovian process a
  walk in a multiparameter space 
  with the consecutive steps labelled with the continuous time variable. 
  The rule governing single steps forward ignores the past history of the walk.
  The iterative solution of the QCD evolution equations can be interpreted
  as a finite Markovian process, limited by the maximum time.
  A Markovian MC implements this process in a natural way.
  In such a MC
  the number of steps is known at the very end of the MC algorithm.}
type, in which the branching process
(the binary decay of the parton) continues until the boundary
of the phase space is hit; the number of branchings (emissions)
is known at the very end of the branching process.
This is in stark contrast to the ME approach, where the number
of partons involved is defined at the very beginning, and the integral 
over standard LIPS is evaluated for a given ME.
In this sense, the ME approach is basically non-Markovian -- this is one of the
(principal) sources of the difficulties in combining the ME and PS approaches.

In this paper we do not offer any ``silver bullet'' solution
of the above problems. 
However, we provide one possibly useful cornerstone,
in constructing yet another class of methods of combining 
ME and PS methodologies.
Our aim is to provide the means of reformulating the PS model 
and the corresponding MC algorithm in a non-Markovian way.
In fact, we restrict ourselves to an even narrower, but well defined
subject of solving QCD DGLAP~\cite{DGLAP} evolution equations
using the MC method, which is at the heart of any PS MC modelling.
We also show, for the initial-state PS (IS PS),
that the {\em non-Markovian} solution
of the DGLAP evolution equations emerges in a natural way as an alternative
solution to yet another long-standing difficulty in the PS MC modelling:
the problem of the {\em energy constraint}.
The energy constraint in the IS PS is the requirement
of constraining to a predefined value the energy of the parton
entering the hard process.
This is so because of a selective nature of the typical hard process ME,
typically due to narrow resonances.
In the typical Markovian PS MC 
the energy of a parton entering the hard process  results 
from many branchings and it is impossible to put any constraint on it,
in the same way as it is impossible to predefine the number of branchings 
or the type of the parton (quark or gluon) at the end of the branching process.
The well known and widely adopted work-around is 
the so-called ``Markovian MC for the backward evolution'' of 
Sj\"ostrand~\cite{Sjostrand:1985xi}%
\footnote{See also ref.~\cite{Marchesini:1988cf}.}.
We shall show that there exists yet another class of MC algorithms 
with the energy constraint, 
which turns out to be non-Markovian in a natural way.

Summarizing, the motivation of our search for non-Markovian modelling
of the QCD evolution equations is that: 
(a) it is closer to the ME approach,
(b) it solves the {\em energy constraint} problem in the IS PS
in a novel way,
with potential advantages of its own, as discussed below.

This paper is one of several related works done in parallel, exploiting
various aspects of the  Markovian-type and non-Markovian-type 
MC solutions of the QCD evolution equations.
Basic results of the present work were presented in the conference
contributions quoted in ref.~\cite{zinnowitz04}.
The earlier work of ref.~\cite{Jadach:2003bu} presents precision MC evaluations
of the LL QCD evolution equations%
\footnote{This work is extended to NLL in ref.~\cite{raport04-08}.}
using an unconstrained Markovian MC.
Although the Markovian calculations of refs.~\cite{Jadach:2003bu,raport04-08}
are not our main aim,
they form a very valuable baseline (benchmark) for the
constrained non-Markovian calculations, as the ones presented here.

This work presents the first successful MC algorithm in 
the constrained non-Markovian
class, although restricted to the pure gluon-strahlung in the
actual numerical implementation.
Later on, the authors of this paper have found
yet another family of MC algorithms,
in the same important class of non-Markovian constrained MCs, 
which will be described in the forthcoming ref.~\cite{raport04-07},
and are even more efficient and easier to implement.
However, at this early stage it makes perfect sense to collect 
{\em all} possible non-Markovian MC algorithms for 
the QCD evolution equations, 
simply because it is difficult to foresee which of them will be most adequate 
in the future attempts at combining PS and ME calculations.
In other words,
the richer the menu of the different non-Markovian algorithms
at our disposal, the better.

The plan of the paper is the following:
in the next section we elaborate more on our aims and the general framework
of our work. 
In section 3 we formulate in detail several examples of the constrained
evolution MC algorithms, and present numerical tests of the corresponding
computer implementations.
A short summary concludes the main result.
The appendix contains the algebra related to the MC method 
(multibranching) employed in section 3.

\section{MC solutions for QCD evolution equations}

As was already said, we are looking for any
possibly non-Markovian,
MC solution of the QCD evolution equations,
with the constraint on the final parton type and its $x$,
the energy fraction.
Needless to say, for a given $x$, 
the solutions of the evolution equations
obtained from the constrained non-Markovian MC 
will be identical to those obtained from
unconstrained MC algorithms, or any other non-MC method --
the real difference is in the efficiency.

The DGLAP evolution equations in QCD, for 
the quark and gluon distributions in the hadron,
are derived in QCD using the renormalization group or diagrammatic
techniques~\cite{DGLAP}.
Let us briefly rederive the {\em iterative solution} 
of these equations. 
We start, as usual, from the evolution equations
in the standard integro-differential form:
\begin{equation}
\begin{split}
\notag
  \label{eq:Evolu}
  \frac{\partial}{\partial t} D_k(t,x)
  &= \sum_j \int\limits_x^1 \frac{d z}{z} P_{kj}(z) 
    \frac{\alpha_S(t)}{\pi} D_j\Big(t,\frac{x}{z} \Big)
= \sum_j {\Peu}_{kj}(t,\cdot)\otimes D_j(t,\cdot),
\end{split}
\end{equation}
where
\begin{displaymath}
f(\cdot){\otimes} g(\cdot)(x) 
    \equiv \int dx_1 dx_2 \delta(x-x_1 x_2)f(x_1)g(x_2)
\end{displaymath}
and 
${\Peu}_{kj}(t,z)\equiv \frac{\alpha_S(t)}{\pi}  P_{kj}(z)$.
Indices $i$ and $k=G,q_a,\bar q_b$ denote gluon, quark and antiquark, while
the evolution { time} is $t=\ln(Q)$.
The differential evolution equation can be turned into the integral equation
\begin{equation}
\begin{split}
\notag
 & e^{\Phi_k(t,t_0)}   D_k(t,x)
    = D_k(t_0,x)
   +\int\limits_{t_0}^t dt_1  e^{\Phi_k(t_1,t_0)}
    \sum_j {\Peu^\Theta_{kj}(t_1,\cdot)}
    \otimes D_j(t_1,\cdot)(x),
\end{split}
\end{equation}
where the IR regulator $\veps$ is introduced:
\begin{eqnarray}
  {\Peu_{kj}(t,z)}&=&-\Peu^{\delta}_{kk}(t,\veps) \delta_{kj}\delta(1-z)
                 +\Peu^{\Theta}_{kj}(t,z),
\\
  \Peu^{\Theta}_{kj}(t,z)&=&\Peu_{kj}(t,z)\Theta(1-z-\veps)
\end{eqnarray}
and the Sudakov form factor
\begin{displaymath}
    {\Phi_{k}(t,t_0)} = \int_{t_0}^t  dt'\;
       \Peu^\delta_{kk}(t',\veps)
\end{displaymath}
appears.
The multiple iteration of the above integral equation leads to:
\begin{equation}
\begin{split}
\notag
 &xD_k(t,x) = e^{-\Phi_k(t,t_0)} xD_k(t_0,x)
  +\sum_{n=1}^\infty \;
   \sum_{k_0...k_{n-1}}
      \bigg[ \prod_{i=1}^n \int\limits_{t_0}^t dt_i\;
      \Theta(t_i-t_{i-1})  \int\limits_0^1 dz_i\bigg]
\\&
      e^{-\Phi_k(t,t_n)}
      \int\limits_0^1 dx_0\;
      \bigg[\prod_{i=1}^n 
           z_i \Peu_{k_ik_{i-1}}^\Theta (t_i,z_i) 
            e^{-\Phi_{k_{i-1}}(t_i,t_{i-1})} \bigg]
     x_0 D_{k_0}(t_0,x_0) \delta\bigg(x- x_0\prod_{i=1}^n z_i \bigg),
  \end{split}
\end{equation}
where $k_n\equiv k$, and the iterative solution is just a series
of integrals ready for integration/simulation with the MC method.
Note that the solution for distributions of parton energies $xD_k(x)$ 
is more convenient, because
kernels obey the energy sum rules: 
\begin{equation}
\label{sumrule}
\sum_l \int dz\; z\Peu_{lk}(z)=0.
\end{equation}
More details can be found in
refs.~\cite{Jadach:2003bu,raport04-08}.

It is well known~\cite{stirling-book} that the above iterative solution can
be implemented as a Markovian process with the probability of every 
single step forward given by the kernel times the Sudakov form factor.
Formal derivation requires adding the extra integration variable 
$t_{n+1},\; t_{n+1}>t,$
in every integral; see ref.~\cite{raport04-08}.
However, for our present purpose the above iterative solution of the evolution 
equations is the proper starting point. 

In ref.~\cite{Jadach:2003bu} it was demonstrated that the 
high-precision Markovian-type MC solution of the evolution equations is 
feasible and it agrees with the non-MC program QCDnum16~\cite{qcdnum16}
to within 0.2\% over a wide range of $x$ and $Q$.

Let us still consider one technical point: the choice of the evolution time.
The MC algorithm will be more efficient if the $t$-dependence
of the strong coupling constant $\alpha_S(t)$ is absorbed by a suitable 
redefinition of the evolution time:
\begin{equation}
\label{eq:tau1}
\tau \equiv \frac{1}{\alpha_S(t_A)} \int_{t_A}^{t} dt_1\; \alpha_S(t_1),\quad
\frac{\partial t}{\partial\tau}= \frac{\alpha_S(t_A)}{\alpha_S(t)}.
\end{equation}
The choice of $t_A$ is arbitrary.
For instance, following the one-loop
$\alpha_S^{(0)}(t)= (2\pi)/( \beta_0 (t-\ln\Lambda_0))$,
we may conveniently choose $t_A$ such that
$\alpha_S^{(0)}(t_A)=2\pi/\beta_0$ 
(e.g. $t_A-\ln\Lambda_0=1$ and hence $t_A=\ln(e\Lambda_0)$).
In such a case $\tau=\ln(t-\ln\Lambda_0)$.
The other choice is $t_A=t_0$, where $t_0$ is the starting point of the evolution.
In either case we have
\begin{equation}
  \label{eq:Iter5}
  \begin{split}
  D_k&(\tau,x) = e^{-\Phi_k(\tau,\tau_0)} D_k(\tau_0,x)
  +\sum_{n=1}^\infty \;
  \int_0^1 dx_0\;
   \sum_{k_0...k_{n-1}}
   \bigg[ \prod_{i=1}^n  \int_{\tau_0}^\tau d\tau_i\; 
     \Theta(\tau_i-\tau_{i-1}) \int_0^1 dz_i\bigg]
\\&~~\times
      e^{-\Phi_k(\tau,\tau_n)}
      \bigg[ \prod_{i=1}^n 
           \Peu_{k_ik_{i-1}}^\Theta (\tau_i,z_i) 
                 e^{-\Phi_{k_{i-1}}(\tau_i,\tau_{i-1})} \bigg]
      D_{k_0}(\tau_0,x_0) \delta\bigg(x- x_0\prod_{i=1}^n z_i \bigg),
  \end{split}
\end{equation}
where $k\equiv k_n$. The kernel $\Peu$ and form factor 
$\Phi_k$ are redefined slightly:
\begin{eqnarray}
  \Peu_{k_ik_{i-1}} (\tau_i,z_i) &= &
   \frac{\alpha_S(t_A)}{\pi}\;
    P_{k_ik_{i-1}} (z_i),
\\
    {\Phi_{k}(\tau,\tau_0)} &=&  \int\limits_{\tau_0}^\tau  d\tau'\;
       \Peu^\delta_{kk}(\veps) 
    = (\tau-\tau_0) 
       \Peu^\delta_{kk}(\veps). 
\end{eqnarray}
In the LL case $\Peu$ is completely independent of $\tau_i$.
In the following we shall usually opt for $t_A=\ln(e\Lambda_0)$)
and $\alpha_S(t_A)/\pi=2/\beta_0$.

\section{Constrained non-Markovian MC algorithms}
\begin{figure}[!ht]
  \centering
  \epsfig{file=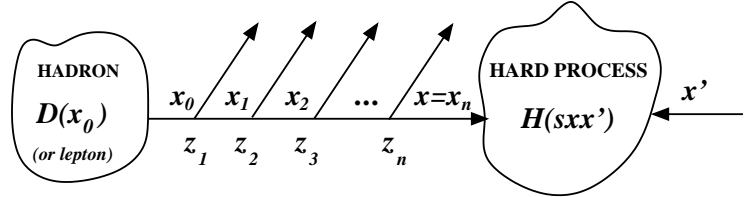,width=100mm}
  \caption{\sf
  Graphical representation of the iterative series of eq.~(\ref{eq:isr1})  
          }
  \label{fig:process}
\end{figure}

\subsection{Solution types I and II}
What are the general classes of the constrained MC solutions? 
Let us write once again the iterative solution of the
evolution equations convoluted with%
\footnote{For simplicity we include here an iterative solution
  of the evolution equations for the single initial-state hadron, but
  our real interest is the case with two initial-state hadrons.}
the parton distribution
$D_{k_0}(\tau_0,x_0)$ at the low energy scale $\tau_0$ and the
hard-process matrix element denoted as $H(x)$
(see also fig.~\ref{fig:process} for the illustration):
\begin{equation}
  \label{eq:isr1}
  \begin{split}
  \sigma  = & \sum_k \int dx H_k(x)D_k(\tau,x)
\\
  =&\sum_k \int dx H_k(x)\;
   \int_0^1 dx_0\;
  \sum_{n=0}^\infty \;
   \sum_{k_{n-1}...k_{1}k_{0}}
   \bigg[ \prod_{i=1}^n\;\; \int\limits_{\tau_{i-1}}^\tau d\tau_i\;
   \int\limits_0^1 dz_i\; \bigg]
\\&~~\times
      e^{-(\tau-\tau_n)R_k}
      \bigg[\prod_{i=1}^n 
      \Peu_{k_ik_{i-1}}^\Theta (z_i)\;
      e^{-(\tau_i-\tau_{i-1})R_{k_{i-1}}} \bigg]\;
      \delta\bigg(x- x_0\prod_{i=1}^n z_i \bigg)\;
      D_{k_0}(\tau_0,x_0),
  \end{split}
\end{equation}
where we define $k\equiv k_n$ and $\prod\limits_{k=1}^0 \equiv 1$
in order to keep the formula compact.
In the LL kernels and form factors simplify
and from eq.~(\ref{sumrule}) it follows that%
\footnote{ In the NLL case an additional $\tau$ dependence through
   $\alpha_S(\tau)$ will invalidate such a simple relation.}: 
\begin{equation}
\label{eq:1formfaktor}
    \Phi_k(\tau,\tau_0)=(\tau-\tau_0)R_k, \qquad
    R_k \equiv \Peu_{kk}^\delta(\veps) =
       \sum_j \int_0^{1-\epsilon} dz\;
        z\Peu^{\Theta}_{jk}(z).
\end{equation}
For the purpose of future discussion we define here additional
virtual form factors:
\begin{equation}
\label{eq:formfaktory}
       R_{jk} = \int_0^{1-\epsilon} dz\;
        z\Peu^{\Theta}_{jk}(z), \quad
    R_k' =
       \sum_{j\neq k} R_{jk} = R_k -R_{kk}.
\end{equation}

Let us discuss basic limitations and
possible solutions for the MC implementation
of the above series of multidimensional integrals.
As already stated, in the ISR case,
since there are narrow resonances in the hard-process function $H(x)$,
the variable $x$ {\em has to be the first one generated} in the MC algorithm,
i.e. it has to be the outermost integration variable.
Similarly it is better to keep $k=k_n$ as the outermost summation
variable as well.

The central issue is the following: How do we treat the variable $x_0$?
There are two possible options.
In the first option (I) $x_0$ is kept as a second outermost integration
variable, next to $x$, i.e. it is generated in the MC as a second variable.
In the second option (II) $x_0$ is treated as one of the last variables in the MC --
in fact it is derived from the other ones using energy constraint.

The following formula describes the first case:
\begin{equation}
  \label{eq:isr2}
  \begin{split}
&  \sigma  =  \sum_{kk_0} \int \frac{dx_0}{x_0} \int dx \;
    H_{k}(x)\; \Deu_{kk_0}\Bigl(\tau,\frac{x}{x_0}\Big|\tau_0\Bigr)\;  
                           D_{k_0}(\tau_0, x_0),
\\
&  \Deu_{kk_0}(\tau,z|\tau_0)
  = \sum_{n=0}^\infty \;
   \sum_{k_{n-1}...k_{1}}
   \bigg[\prod_{i=1}^n\;\; \int\limits_{\tau_{i-1}}^\tau d\tau_i\;
   \int\limits_0^1 dz_i\;\bigg]
\\&~~~~~~\times
      e^{-(\tau-\tau_n)R_k}
      \bigg[ \prod_{i=1}^n 
      \Peu_{k_ik_{i-1}}^\Theta (z_i)\;
      e^{-(\tau_i-\tau_{i-1})R_{k_{i-1}}} \bigg]\;
      \delta\bigg(z- \prod_{i=1}^n z_i \bigg),
  \end{split}
\end{equation}
with $k\equiv k_n$ as usual.
We shall refer to this option as to a {\em solution type I};
this is the scenario that was implemented for the QED ISR 
pure brems\-strahlung in several YFS-type MC programs, 
starting from the prototype of ref.~\cite{Jadach-yfs-mpi:1987}.
The main technical difficulty is the implementation/elimination
of the $\delta(z-\prod z_i)$ function. 
This problem was solved in QED by eliminating the delta function with
integration over the $z$ of the hardest photon%
\footnote{ In this QED case the integration over $x_0$ is rather
  trivial because one starts from $D(\tau_0,z)=\delta(1-z)$.}
(the largest $1-z$).
This scenario looks definitely feasible, %
and the first working example will be described in a separate work; 
see ref.~\cite{raport04-07}.

The second scenario, referred to as {\em solution type II}
relies on the fact that the starting parton distribution for typical hadron
beam particle $D_{k_0}(\tau_0, x)$
can be relatively well (by MC standards) approximated
by a power-like function $D_{k_0}(\tau_0, x) \sim x^{-1+\eta}$
over a wide range of $\ln(x)$ (i.e.\ $10^{-4}<x<1$),
with the parameter $\eta$ not far from zero.
In fact the gluon and quark singlet parton distributions of the nucleon at 
low $Q$ feature %
$\eta \simeq -0.2$.
In solutions of type II
the essential idea is that, in eq.~(\ref{eq:isr1}), 
the $\delta\big(x- x_0\prod_{i=1}^n z_i \big)$
is eliminated  using the integration over $x_0$. 
It means that $x_0$, contrary to type I,
is generated in the MC as a last variable instead of a second one.
More precisely it is not generated at all, but determined as a function
of all previously generated variables $x_0=x\bigl(\prod z_i\bigr)^{-1}$,
by means of solving the energy constraint.

Let us isolate explicitly the small-$z$ limit
from the starting parton distribution:
\begin{equation}
  D_{k_0}(\tau_0,x_0) = W^{*D}_{k_0}(x_0)\; A_{k_0} x_0^{-1+\eta},\qquad
  W^{*D}_{k_0}(x_0) 
  \equiv \frac{D_{k_0}(\tau_0,x_0)}{A_{k_0} x_0^{-1+\eta}} \leq 1,
\end{equation}
where $W^{*D}_k$ is the MC weight to be neglected now and restored later on.
Elimination of the $\delta$-functions with the help of the $x_0$-integration
leads to
\begin{equation}
  \label{eq:sc2}
  \begin{split}
  \sigma
  =&\sum_k \int dx\; H_k(x)\; 
   A_{k_0} x^{-1+\eta}\;
  \sum_{n=0}^\infty \;
   \sum_{k_{n-1}...k_{1}k_{0}}
   \bigg[\prod_{i=1}^n\;\; \int\limits_{\tau_{i-1}}^\tau d\tau_i\;
                           \int\limits_0^1 dz_i\; \bigg]
\\&~~\times
      e^{-(\tau-\tau_n)R_k}
      \bigg[ \prod_{i=1}^n  z_i^{-\eta}
      \Peu_{k_ik_{i-1}}^\Theta (z_i)\;
      e^{-(\tau_i-\tau_{i-1})R_{k_{i-1}}} \bigg]\;
      \Theta \Big(\prod_{j=1}^n z_j -x \Big)
      W^{*D}_{k_0}(\bar x_0),
  \end{split}
\end{equation}
where $\bar x_0=x/\prod_j z_j$.
For the distributions of quarks and gluons in the proton at a low energy scale 
we have the same $\eta\simeq -0.2$. 
Hence, we may also include all factors $z_i^{-\eta}$ in the
MC weight $W^{D}_k\leq 1$, to be neglected and restored at a later stage
of the MC algorithm together with the other
details of the $D_k(\tau_0,x_0)$ function, or
in a more sophisticated MC algorithm, we may actually generate
exactly the distributions $z_i^{-\eta} \Peu_{k_ik_{i-1}}^\Theta (z_i)$.
In the following let us assume the former simpler case 
with $z_i^{-\eta}$ in the MC weight:
\begin{equation}
  \label{eq:sc5}
  \begin{split}
  &\sigma
  =\sum_k \int dx\; 
   H_k(x)\; A_{k_0} x^{-1+\eta}\;
  \sum_{n=0}^\infty \;
   \sum_{k_{n-1}...k_{1}k_{0}}
   \bigg[ \prod_{i=1}^n\;\; \int\limits_{\tau_{i-1}}^\tau d\tau_i\;
                            \int\limits_0^1 dz_i\; \bigg]
\\&~~~~~~~\times
      e^{-(\tau-\tau_n)R_k}
      \bigg[ \prod_{i=1}^n  
      \Peu_{k_ik_{i-1}}^\Theta (z_i)\;
      e^{-(\tau_i-\tau_{i-1})R_{k_{i-1}}} \bigg]\;
      \Theta \Big(\prod_{j=1}^n z_j -x\Big)
      W^{D}_k(x,\bar x_0),
\\&
   W^{D}_{k_0}(x,\bar x_0) 
   = \frac{D_{k_0}(\tau_0,\bar x_0)}{A_{k_0} \bar x_0^{-1+\eta}} 
   \bigg\{\prod_{i=1}^n z_i \bigg\}^{-\eta} 
   = \frac{D_{k_0}(\tau_0,\bar x_0)}{A_{k_0} \bar x_0^{-1+\eta}} 
   \bigg\{ \frac{x}{\bar{x}_0} \bigg\}^{-\eta} 
   \leq 1,
  \end{split}
\end{equation}
where $D_{k_0}(\tau_0,\bar x_0) \to A_{k_0} \bar x_0^{-1+\eta}$ 
in the limit $\bar x_0 \to 0$.

Let us stress that eq.~(\ref{eq:sc2}) implements the exact iterative
distribution of the evolution equations.
In the following sections
we will show the results from the prototype MC
based directly on the above expression for the pure bremsstrahlung case.
As we shall see, it works well for the case of the emission from a quark;
however, it has rather low acceptance  ($\sim 10^{-3}$) for the emission
from a gluon.
We shall call such a MC solution, directly based on eq.~(\ref{eq:sc2}),
solution type II.a.

In another method, II.b, we shall reorganize the integration variables
in a hierarchical way, 
and use multibranching to isolate the $1/z$ part of the gluon-to-gluon kernel.
Such a solution is rather complicated and non-trivial to implement in a
general case 
with multiple flavour-changing (quark--gluon) transitions.
However, successful implementation of the pure gluon-strahlung case, 
presented in the next section,
allows us to claim that the efficiency of the MC type II.b is satisfactory,
and the gate to  practical applications of MC solutions of this type 
is wide open.

\subsection{Constrained Monte Carlo: solutions class II}
As already noticed in ref.~\cite{Jadach:2003bu},
in the long emission chain (on average $\sim 20$ emissions), 
from $Q=1$ GeV to $Q=1$ TeV,
most of the emissions are of bremsstrahlung type, i.e.
they preserve the identity of the parton on the main line of the chain.
It was shown there that  on the average
only about one out of twenty emissions involves
flavour transmutation, $q\to G$ or $G\to q$; 
the other ones are gluon emissions.
With this in mind, it is therefore natural
to reorganize the iterative solution of the evolution equations in such
a way that
all pure bremsstrahlung adjacent vertices in the emission chain are 
lumped together
into segments described by the following universal evolution function:
\begin{equation}
  \label{eq:dfun}
  \begin{split}
  d'_k(\tau,z | \tau_0) 
  =& e^{-(\tau-\tau_0)R_{kk}}
   \sum_{n=0}^\infty\;
   \prod_{i=1}^n\; \int\limits_{\tau_{i-1}}^\tau d\tau_i\;
                   \int\limits_0^1 dz_i\;
      \Peu_{kk}^\Theta (z_i)\;
      \delta\bigg(z- \prod_{j=1}^n z_j \bigg),
  \end{split}
\end{equation}
where the type of parton on the main line, $k=Q$ or $k=G$, is unchanged.
Note that we have retained in eq.~(\ref{eq:dfun}) only part of the
virtual form factor $R_k$, namely the $R_{kk}$ function, 
which matches exactly the
real emission kernel $\Peu_{kk}^\Theta$ of pure bremsstrahlung. 
The leftover $R'_k$ is included explicitly in the following
eq.~(\ref{eq:hiera2}).

The full iterative solution of the previous section can be expressed
in terms of the product of the above functions and the kernels representing
flavour-changing transitions $q\to G$ or $G\to q$ in the following way:
\begin{equation}
  \label{eq:hiera2}
  \begin{split}
  D_k(\tau,x)
 &=\sum_{n=0}^\infty \;
   \int\limits_0^1 dx_0\;
   \sum_{{k_{n-1}\dots,k_{1},k_{0}}\atop %
         {k_{j}\neq k_{j-1}, j=1,\dots,n}}
   \Biggl(\prod_{i=1}^n\; \int\limits_{\tau_{i-1}}^\tau d\tau_i\Biggr)
   \Biggl(\prod_{i=1}^n \int\limits_0^1 dz_i\Biggr)
   \Biggl(\prod_{i=1}^{n+1} \int\limits_0^1 dz'_i\Biggr)
\\&~~~\times
      e^{-(\tau-\tau_n)R'_k}\;
      d'_{k}(\tau,z'_{n+1} | \tau_{n})\;
\\&~~~\times
      \prod_{i=1}^n \bigg[
      \Peu_{k_ik_{i-1}}^\Theta (z_i)\;
      e^{-(\tau_i-\tau_{i-1})R'_{k_{i-1}}}\;
      d'_{k_{i-1}}(\tau_i,z'_i | \tau_{i-1}) \bigg]\;
\\&~~~\times
      D_{k_0}(\tau_0,x_0) 
      \delta\bigg(x- x_0 \prod_{i=1}^n z_i \prod_{i=1}^{n+1} z'_i \bigg),
  \end{split}
\end{equation}
where we employ the usual conventions: 
$k_n\equiv k$ and $\prod_{j=1}^0\equiv 1$.

We say that the above formula implements {\em hierarchical} organization 
of the emission chain,
because it represents the Markovian process in which each pure bremsstrahlung
step (segment) in the Markovian random walk (emission chain)
is an independent Markovian process of its own!
This elegant and powerful reorganization
of the emission chain is proved formally in the separate work of 
ref.~\cite{raport04-09}. 

The complete hierarchical formula for the integrated cross section,
after elimination of the delta function with $x_0$ integration,
reads as follows:
\begin{equation}
  \label{eq:hiera3}
  \begin{split}
  \sigma
  =\sum_k \int dx\; & H_k(x)\; 
   A_{k_0} x^{-1+\eta}\;
   \sum_{n=0}^\infty \;
   \sum_{{k_{n-1}\dots,k_{1},k_{0}}\atop %
         {k_{j}\neq k_{j-1}, j=1,\dots,n}}
   \prod_{i=1}^n\; \int\limits_{\tau_{i-1}}^\tau d\tau_i\;
   \prod_{i=1}^n \int\limits_0^1 dz_i\; 
   \prod_{i=1}^{n+1} \int\limits_0^1 dz'_i\;
\\&~~~\times
      e^{-(\tau-\tau_n)R'_k}\;
      d'_{k}(\tau,z'_{n+1} | \tau_{n})\;
\\&~~~\times
      \prod_{i=1}^n \bigg[
      \Peu_{k_ik_{i-1}}^\Theta (z_i)\;
      e^{-(\tau_i-\tau_{i-1})R'_{k_{i-1}}}\;
      d'_{k_{i-1}}(\tau_i,z'_i | \tau_{i-1}) \bigg]\;
\\&~~~\times
      \Theta\bigg(\prod_{i=1}^n z_i \prod_{i=1}^{n+1} z'_i - x \bigg)
      W^D_{k_0}(x,\bar x_0),
\\
  d'_k(\tau_i,z'_i | \tau_{i-1}) 
  =& e^{-(\tau_i-\tau_{i-1})R_{kk}}
   \sum_{n^{(i)}=0}^\infty\;
   \prod_{m=1}^{n^{(i)}}\; 
       \int\limits_{\tau^{(i)}_{m-1}}^{\tau^{(i)}} d\tau^{(i)}_m\;
   \int\limits_0^1 dz^{(i)}_m\;
      \Peu_{kk}^\Theta (z^{(i)}_m)\;
      \delta\bigg(z'_i- \prod_{m=1}^{n^{(i)}} z^{(i)}_m \bigg).
  \end{split}
\end{equation}
where
\begin{equation}
\tau^{(i)}\equiv\tau_i,\quad
\tau^{(i)}_0\equiv\tau_{i-1},\quad
\bar x_0=\frac{x}{\prod_{i=1}^n z_i \prod_{i=1}^{n+1} z'_i } \leq 1.
\end{equation}
The above formula is the starting point in the next two subsections
for construction of constrained MC algorithms of type II.

\subsubsection{Solution II.a}
In the straightforward solution of type II, which we call II.a,
the single $\Theta$-function for both
flavour-changing emissions and pure brems\-strahlung segments is replaced
by the product of the individual $\Theta$-functions --
thus decoupling completely the $z$-integration space
and opening the way for
the analytical integrations of the approximate spectra
for the purpose of the MC generation.
More precisely, let us first notice that
we are really dealing with the single $\Theta$-function 
involving {\em all} $z$ variables due to trivial identity
\begin{equation}
  \begin{split}
  \prod_{i=1}^{n+1} \Biggl[
  \int\limits_0^1 dz'_i\; \biggl(
  \prod_{m=1}^{n^{(i)}}\; \int\limits_0^1 dz^{(i)}_m\;
                          \biggr)
      \delta\bigg(z'_i- \prod_{m=1}^{n^{(i)}} z^{(i)}_m \bigg)
                     \Biggr]
 &\Theta\bigg(\prod_{i=1}^n z_i \prod_{i=1}^{n+1} z'_i -x\bigg)=
\\
=  
  \prod_{i=1}^{n+1}      \biggl(
  \prod_{m=1}^{n^{(i)}}\; \int\limits_0^1 dz^{(i)}_m\;
                          \biggr)
 &\Theta\bigg(
    \prod_{i=1}^n z_i 
    \prod_{i=1}^{n+1} 
      \prod_{m=1}^{n^{(i)}} z^{(i)}_m -x \bigg).
  \end{split}
\end{equation}

We are now ready to describe the essence of the MC algorithm of type II.a.
We use the following identity
\begin{equation}
  \begin{split}
 &\Theta\bigg(
    \prod_{i=1}^n z_i 
    \prod_{i=1}^{n+1} 
      \prod_{m=1}^{n^{(i)}} z^{(i)}_m -x \bigg)
= W^{\Theta}_{II.a}\;\;
 \prod_{i=1}^n \Theta(z_i -x)
 \prod_{i=1}^{n+1} \prod_{m=1}^{n^{(i)}}  
  \Theta\bigg(z^{(i)}_m -x \bigg),
  \end{split}
\end{equation}
where the function
\begin{equation}
  \begin{split}
& W^{\Theta}_{II.a} \equiv
    \Theta\left(
    \prod_{i=1}^n z_i \prod_{i=1}^{n+1} 
      \prod_{m=1}^{n^{(i)}} z^{(i)}_m -x\right)
    \leq 1
  \end{split}
\end{equation}
is the Monte Carlo weight.
This MC weight will be neglected later on, so that variables
can be generated according to simplified
distributions, and finally the generated events will be weighted 
according to this weight.

Let us rewrite our master integral (\ref{eq:hiera3}) without any 
approximations
\begin{equation}
  \begin{split}
  \sigma
  &=\sum_k \int dx\;  H_k(x)\;
   A_{k_0} x^{-1+\eta}\;
   \sum_{n=0}^\infty \;
   \sum_{{k_{n-1}\dots,k_{1},k_{0}}\atop %
         {k_{j}\neq k_{j-1}, j=1,\dots,n}}
   \bigg[ \prod_{i=1}^n\; \int\limits_{\tau_{i-1}}^\tau d\tau_i\;
                          \int\limits_x^1 dz_i\;\bigg]
\\&\times
      e^{-(\tau-\tau_n)R'_k}
      d''_{k}(\tau,x | \tau_{n})
      \bigg[ \prod_{i=1}^n 
      \Peu_{k_ik_{i-1}}^\Theta (z_i)
      e^{-(\tau_i-\tau_{i-1})R'_{k_{i-1}}}
      d''_{k_{i-1}}(\tau_i,x | \tau_{i-1}) \bigg]
      W^D_{k_0}(x,\bar x_0) W^{\Theta}_{\rm II.a},
\\
  d&''_k(\tau_i,x| \tau_{i-1}) 
  = e^{-(\tau_i-\tau_{i-1})R_{kk}}
   \sum_{n^{(i)}=0}^\infty\;
   \prod_{m=1}^{n^{(i)}}\; 
       \int\limits_{\tau^{(i)}_{m-1}}^{\tau^{(i)}} d\tau^{(i)}_m\;
   \prod_{m=1}^{n^{(i)}}\; \int\limits_x^1 dz^{(i)}_m\;
      \Peu_{kk}^\Theta (z^{(i)}_m)
  \end{split}
\end{equation}
where
$\bar x_0 =x/\Big( \prod_{i=1}^n z_i \prod_{i=1}^{n+1} 
  \prod_{m=1}^{n^{(i)}} z^{(i)}_m \Big)$ and 
one should remember that $d''$-functions provide
integration over all the $z^{(i)}_i$ variables that are implicitly present
in the function $W^{\Theta}_{\rm II.a}$.
The enormous advantage of the above procedure is that in the approximate
integral we can sum up and integrate immediately over all brems\-strahlung
segments of the emission chain.
To see it let us drop the two MC weights $W^D_{k_0}$ and
$W^{\Theta}_{\rm II.a}$. 
Now we can immediately sum up and integrate analytically the pure 
brems\-strahlung subintegrals:
\begin{equation}
  d''_k(\tau_i,x| \tau_{i-1}) 
  = \exp\biggl(
       (\tau_i-\tau_{i-1}) \Bigl(-R_{kk}
              +\int\limits_x^1 dz\;
      \Peu_{kk}^\Theta (z)\Bigr)
        \biggr)
\end{equation}
and hence
\begin{equation}
  \begin{split}
  \sigma
  &=\sum_k \int dx\;  H_k(x)\; 
   A_{k_0} x^{-1+\eta}\;
   \sum_{n=0}^\infty \;
   \sum_{{k_{n-1}\dots,k_{1},k_{0}}\atop %
         {k_{j}\neq k_{j-1}, j=1,\dots,n}}
   \bigg[ \prod_{i=1}^n\; \int\limits_{\tau_{i-1}}^\tau d\tau_i\;
                          \int\limits_x^1 dz_i\; \bigg]
\\&\times
      e^{(\tau-\tau_n) (- R_k +\int_x^1 dz \Peu_{kk}^\Theta (z)) }\;
      \bigg[ \prod_{i=1}^n 
      \Peu_{k_ik_{i-1}}^\Theta (z_i)\;
      e^{(\tau_i-\tau_{i-1}) ( -R_{k_{i-1}} 
               +\int_x^1 dz \Peu_{k_{i-1}k_{i-1}}^\Theta (z))} \bigg].
\end{split}
\end{equation}
The above looks rather promising, because  we are
left with the relatively simple problem of generating
several variables $\tau_i$ and $z_i$ ($i\leq 5$ seems sufficient)
for the flavour-changing emissions, for which the above integrals
provide explicit analytical distribution.
The MC events are attributed with the MC weight 
$W_{\rm II.a}= W^D_{k_0}(\bar x_0)\; W^{\Theta}_{\rm II.a} \leq 1$.
The key question is: What is the acceptance rate for this weight?
We did an introductory exercise, implementing the pure brems\-strahlung
version of it%
.
We have found that, unfortunately, the
acceptance rate for the emission from the gluon line is only about $10^{-4}$.
This inefficiency can be traced back
to the presence of the $1/z$ singularity in the $G\to G$ kernel.
Namely, if we allow for the range
$x>x_{\min}=10^{-5}$ we also allow for $z$ to be generated to the same
low limit. In the case of $\Peu_{GG}(z)$ containing  a $1/z$ part,
this creates many events with low $z_i$.
Consequently, $\bar x_0 =x/\prod z_i$ goes very often beyond 1 and
the corresponding MC weight $W^{\Theta}_{\rm II.a}$ gets zero value.

On the other hand, for the quark line the acceptance rate
is close to 1, which is clearly related to the absence of
the component $1/z$ in $\Peu_{qq}(z)$.

The above numerical exercise indicates that the 
$1/z$ part in the $\Peu_{GG}(z)$ has to be treated better, as is done
in the present case II.a.
A more sophisticated treatment of the $1/z$ component of the kernel
is applied in the solution II.b described in the next section.

The present solution II.a is still a workable solution. 
In spite of its very low efficiency, 
due to its relative simplicity, it can still be 
quite useful for testing other more sophisticated solutions.
We therefore implemented it also in the MC program, for the moment
only in the pure bremsstrahlung version.

\subsubsection{Solution II.b}
\label{sect:II.b}
In this section we present a solution 
more sophisticated than the II.a one of the previous section:
we split the brems\-strahlung kernel $\Peu_{GG}$
into two parts: (B) $\sim 1/z$ and (A) = the rest.
Then we apply the multibranching%
\footnote{Multibranching or multichannelling is the standard MC technique
  in which the distribution is split into a sum of positively 
  defined subdistributions.
  First the index numbering the distributions is generated.
  Once the subdistribution is chosen, a
  MC point is generated according to this subdistribution, 
  instead of the total distribution; 
  see ref.~\cite{mcguide:1999} for details.},
as described in Appendix~A,
to every pure brems\-strahlung segment of the emission chain.
Finally, we also treat the $\Theta$-function more selectively than in II.a.
The product of the individual $\Theta$-functions appears for
the flavour-changing emissions and part (A) of the brems\-strahlung kernel,
while the single $\Theta$-function is left for each segment describing
part (B) of the pure brems\-strahlung.
Such partial decoupling in the $z$-integration space still allows
for the analytical integration of the approximate spectra
for the purpose of the early stage of the MC generation.
This is possible because segments of type (B) in the pure brems\-strahlung
parts are integrable (to a Bessel function), as shown below,
while for the rest we get exponentials in a way similar to those in II.a.

Again, the starting point is
the complete hierarchical formula (\ref{eq:hiera3}) 
for the integrated cross section.
In the case of the gluonic subintegral $d'_{k=G}(\tau_i,z'_i|\tau_{i-1})$
we reorganize this integral to isolate
the $1/z$ part from the kernel.
In order to split the $\Peu_{GG}(z)$
in two {\em positive} parts, one of them being $1/z$,
we have to simplify it first:
\begin{equation}
  \begin{split}
  &\Peu^\Theta_{GG}(z) = \bar\Peu^\Theta_{GG}(z)\; w_{GG}(z),\quad
   P^\Theta_{GG}(z)    = \bar P^\Theta_{GG}(z)\; w_{GG}(z),\quad
\\&
   \bar P^\Theta_{GG}(z) = 2C_A \left( \frac{1}{z}+\frac{1}{1-z}\right),\quad
   w_{GG}(z) = (1-z(1-z))^2 \leq 1.
  \end{split}
\end{equation}
Having done that and using the multibranching identity of eq.~(\ref{eq:identity})
in the appendix
we may rewrite the gluonic brems\-strahlung subintegral for $k=G$ as follows:
\begin{equation}
  \begin{split}
 &d'_{k}(\tau_i,z'_i | \tau_{i-1}) 
  = \int\limits_0^1 \frac{dZ^{(i)}}{Z^{(i)}}\;\Theta(Z^{(i)}-z'_i)
  d^B_k(\tau_i, Z^{(i)}|\tau_{i-1})\,
  d^A_k\biggl(\tau_i, \frac{z'_i}{Z^{(i)}}|\tau_{i-1}\biggr)
   W_{kk}\bigl({\bf z^{(i)}}({\bf z^{\bu(i)}, {z'}^{(i)}} )\bigr),
  \end{split}
\end{equation}
where
\begin{equation}
  \begin{split}
&
  d^A_k(\tau_i, Z^{(i)}|\tau_{i-1})
= e^{-(\tau_i-\tau_{i-1})R^A_{kk}}
\\&~~\times 
  \sum_{n'^{(i)}=0}^\infty\;\;
     \prod_{m=1}^{n'^{(i)}}
        \int\limits^{\tau_i}_{\tau_{i-1}} d\tau'^{(i)}_m \;
        \Theta(\tau'^{(i)}_m - \tau'^{(i)}_{m-1})
     \prod_{m=1}^{n'^{(i)}} 
        \int dz_m'^{(i)}\;
        \bar\Peu_{kk}^{\Theta A} (z_m'^{(i)})\;
     \delta\bigg(Z^{(i)} -\prod_{m=1}^{n'^{(i)}} z_m'^{(i)} \bigg)
\\
&
  d^B_k(\tau_i, Z^{(i)}|\tau_{i-1})
= e^{-(\tau_i-\tau_{i-1})R^B_{kk}}
\\&~~\times 
  \sum_{n^{\bu(i)}=0}^\infty\;\;
     \prod_{m=1}^{n^{\bu(i)}}
        \int\limits^{\tau_i}_{\tau_{i-1}} d\tau^{\bu(i)}_m \;
        \Theta(\tau^{\bu(i)}_m - \tau^{\bu(i)}_{m-1})
     \prod_{m=1}^{n^{\bu(i)}} 
        \int dz_m^{\bu(i)}\;
        \bar\Peu_{kk}^{\Theta B} (z_m^{\bu(i)})\;
     \delta\bigg(Z^{(i)} -\prod_{m=1}^{n^{\bu(i)}} z_m^{\bu(i)} \bigg)
  \end{split}
\end{equation}
and
\begin{equation}
  \begin{split}
	&\bar P^{\Theta A}_{GG}(z) = \frac{2C_A}{1-z},\qquad
	 \bar P^{\Theta B}_{GG}(z) = \frac{2C_A}{z}, \Biggl.
\\
  &\Peu^{\Theta B}_{GG}(z) = \bar\Peu^{\Theta B}_{GG}(z),\qquad
   \Peu^{\Theta A}_{GG}(z) = \Peu^\Theta_{GG}(z) -\Peu^{\Theta B}_{GG}(z).
  \end{split}
\end{equation}
In the above we used the notation
${\bf {z'}^{(i)}} =
     \bigl({z'}^{(i)}_1,\dots,{z'}^{(i)}_{n'^{(i)}}\bigr)$ 
and ${\bf z^{\bu(i)}} =
     \bigl({z}^{\bu(i)}_1,\dots,{z}^{\bu(i)}_{n^{\bu(i)}}\bigr)$.

The MC weight due to the kernel simplification
\begin{equation}
   W_{GG}\bigl({\bf z^{(i)}}({\bf z^{\bu(i)}, {z'}^{(i)}} )\bigr)
     = \prod_{m=1}^{n^{\bu(i)} + {n'}^{(i)}} 
      w_{GG}\bigl(z^{(i)}_m({\bf z^{\bu(i)}, {z'}^{(i)}} )\bigr),
\end{equation}
depends on the variables $z^{(i)}_m$ after {\em relabelling}.
A special kind of permutation
$z^{(i)}_m\to z^{\bu(i)}_l, {z'}^{(i)}_l$,
which we refer to as {\em relabelling}, is 
an important part of the MC algorithm --
it is defined precisely in the appendix.
Since relabelling is just a permutation of $z$'s, we may calculate
the weight $W_{GG}$ with the $z^{\bu(i)}_l, {z'}^{(i)}_l$
variables before the relabelling:
\begin{equation}
     W_{GG}=
       \prod_{m=1}^{n^{\bu(i)}} w_{GG}(z^{\bu(i)}_m)\;\;
       \prod_{m=1}^{{n'}^{(i)}} w_{GG}({z'}^{(i)}_m).
\end{equation}

Until now we made no approximation in our master integral --
we only reorganized integration variables, in particular isolating
the $1/z$ component in the pure brems\-strahlung subintegrals for the
gluon emitters.
In full analogy to case II.a,
in the last step in this reorganization we eliminate all variables $z'_i$
and a class of $\delta$-functions 
with the help of the identity:
\begin{equation}
  \begin{split}
  \prod_{i=1}^{n+1} \int\limits_0^1 dz'_i\;\Theta(Z^{(i)}-z'_i)\;
 &
    \prod_{i=1}^{n+1} \delta\bigg(
               \frac{z'_i}{Z^{(i)}}
             - \prod_{m=1}^{{n'}^{(i)}} {z'}^{(i)}_m \bigg)~
  \Theta\bigg(\prod_{i=1}^n z_i \prod_{i=1}^{n+1} z'_i -x \bigg)=
\\
=&  
 \prod_{i=1}^{n+1} Z^{(i)}\;
 \Theta\Bigg( 
    \prod_{i=1}^n z_i
    \prod_{i=1}^{n+1} \bigg(Z^{(i)} 
      \prod_{m=1}^{{n'}^{(i)}} {z'}^{(i)}_m\bigg) 
    -x\Bigg).
  \end{split}
\end{equation}
Consequently, from now on we substitute $z'_i$ with $\bar z'_i$
\begin{equation}
z'_i \to \bar z'_i= Z^{(i)} \prod_{m=1}^{{n'}^{(i)}} {z'}^{(i)}_m \leq 1.
\end{equation}
On the other hand, we keep $\delta$-functions inside the $d_G^B$ functions,
which will also be treated analytically, but separately; see below.

Now comes the essential step in the algorithm II.b 
-- we define the following MC weight:
\begin{equation}
  \begin{split}
& \Theta\bigg( 
    \bigg\{
    \prod_{j=1}^n z_j 
    \prod_{j=1}^{n+1}\;  Z^{(j)} \bigg\}
    \bigg\{
    \prod_{i=1}^{n+1}\;  \prod_{m=1}^{{n'}^{(i)}} 
           {z'}^{(i)}_m \bigg\} 
   -x \bigg)=
\\&~~~~~~~~~~~~~~~~~
= W^{\Theta}_{\rm II.b}~
 \Theta\bigg(
     \prod_{j=1}^n z_j \prod_{j=1}^{n+1}\;  Z^{(j)}
    -x \bigg)
 \prod_{i=1}^{n+1} 
 \prod_{m=1}^{{n'}^{(i)}}  
  \Theta\bigg( 
    \bigg\{
    \prod_{j=1}^n z_j 
    \prod_{j=1}^{n+1}\;  Z^{(j)} \bigg\}
           {z'}^{(i)}_m 
     -x  \bigg)
\\&~~~~~~~~~~~~~~~~~
= W^{\Theta}_{\rm II.b}~
 \Theta\Big(1 - z^{\rm eff}_{\min}({\bf z},{\bf Z}) \Big)
 \prod_{i=1}^{n+1} 
 \prod_{m=1}^{{n'}^{(i)}}  
   \Theta\Big({z'}^{(i)}_m -z^{\rm eff}_{\min}({\bf z},{\bf Z})\Big),
  \end{split}
\end{equation}
where
\begin{equation}
  z^{\rm eff}_{\min}\Big(z_1,z_2,...,z_n,Z^{(1)},Z^{(2)},...,Z^{(n+1)}\Big)\; 
     \equiv\;
     \frac{x}{\prod_{j=1}^n z_j \prod_{j=1}^{n+1}\;  Z^{(j)} }.
\end{equation}

The new MC weight 
\begin{equation}
  \begin{split}
& W^{\Theta}_{\rm II.b} \equiv
  \Theta\left(
    \prod_{i=1}^n z_i 
    \prod_{i=1}^{n+1} \biggl(Z^{(i)} 
      \prod_{m=1}^{{n'}^{(i)}} z^{(i)}_m\biggr) 
          -x \right)
  \leq 1
  \end{split}
\end{equation}
will be neglected later on
and restored at the end as a standard MC compensating weight%
\footnote{There are a few other slightly different possible choices
  of $z^{\rm eff}_{\min}$, which are not discussed here.}.

All these preparatory steps lead us to the following 
{\em master equation for method II.b},
still without any approximation, but with clearly defined
MC weights and the distributions to be generated at the early stage of the
MC algorithm:
\begin{equation}
  \label{eq:IIBmaster}
  \begin{split}
  \sigma
  =\sum_k \int dx\; &  H_k(x)\; 
   A_{k_0} x^{-1+\eta}\;
   \sum_{n=0}^\infty \;
   \sum_{{k_{n-1}\dots,k_{1},k_{0}}\atop %
         {k_{j}\neq k_{j-1}, j=1,\dots,n}}
   \prod_{i=1}^n\; \int\limits_{\tau_{i-1}}^\tau d\tau_i\;
   \prod_{i=1}^n \int\limits_0^1 dz_i\; 
   \prod_{i=1}^{n+1} \int\limits_0^1 \frac{dZ^{(i)}}{Z^{(i)}}\;
\\&\times
      e^{-(\tau-\tau_n)R'_k}\;
      {d^A_{k}}'(\tau,z^{\rm eff}_{\min} | \tau_{n})\;
      d^B_k(\tau, Z^{(n+1)}|\tau_{n})
\\&\times
      \prod_{i=1}^n \bigg[
      \Peu_{k_ik_{i-1}}^\Theta (z_i)\;
      e^{-(\tau_i-\tau_{i-1})R'_{k_{i-1}}}\;
      {d^{A}_{k_{i-1}}\hskip -3mm}'\hskip 3mm
                    (\tau_i,z^{\rm eff}_{\min} | \tau_{i-1})\;
      d^B_k(\tau_i, Z^{(i)}|\tau_{i-1})
      \bigg]\;\;
\\&\times
      W^D_{k_0}(\bar x_0)\;
      W^{\Theta}_{\rm II.b}\;
      {\Theta(1 -z^{\rm eff}_{\min})},
\\
  {d^A_{k}}'(\tau_i,z^{\rm eff}_{\min} | \tau_{i-1}) 
 &= e^{-(\tau_i-\tau_{i-1})R^A_{kk}}
   \sum_{n'^{(i)}=0}^\infty\;
   \prod_{m=1}^{n'^{(i)}}\; 
       \int\limits_{\tau'^{(i)}_{m-1}}^{\tau_i} d\tau'^{(i)}_m\;
\\&\times
   \prod_{m=1}^{n'^{(i)}}\; \int\limits_0^1 dz'^{(i)}_m\;
      \bar\Peu_{kk}^{\Theta A} (z'^{(i)}_m)\;
      \Theta\big(z'^{(i)}_m - z^{\rm eff}_{\min} \big)~
      W_{GG}({\bf z^{\bu{(i)}}},{\bf {z'}^{(i)}}).
  \end{split}
\end{equation}
The functions $d^B_k(\tau_i, Z^{(i)}|\tau_{i-1})$ and the variables $Z^{(i)}$
are really present only for gluon, $k=G$. However,
in order to keep the notation compact, we understand that
\begin{equation}
d^B_{k\neq G}(\tau_i, Z^{(i)}|\tau_{i-1}) \equiv \delta(1-Z^{(i)}).
\end{equation}
We also assume $\prod_{n=1}^0=1$, as usual.
The reader should also keep in mind that at this stage the
integrand of the part ${d^A}'$ still
depends on the integration variables of $d^B_G$.

In the MC algorithm of type II.b all three MC weights are neglected:
\begin{equation}
      W^D_{k_0}(\bar x_0)\;
      W^{\Theta}_{\rm II.b},
      W_{GG}({\bf z^{\bu{(i)}}},{\bf {z'}^{(i)}})
  \equiv W_{\rm II.b}
\end{equation}
at the early stage of the MC algorithm and later on
generated MC events are weighted with $W_{\rm II.b}$.
Since $W_{\rm II.b} \leq 1$, we may easily transform weighted MC events
into unweighted events by rejecting some of the MC events in the usual way.

The MC weight $W_{\rm II.b}$
was chosen in such a way that once it is neglected, we can
perform a lot of analytical integrations:
\begin{equation}
  \label{eq:II.bmc}
  \begin{split}
  \sigma_{\hbox{\footnotesize prim.}}
  =\sum_k \int & dx\;  H_k(x)\; 
   A_{k_0} x^{-1+\eta}\;
   \sum_{n=0}^\infty \;
   \sum_{{k_{n-1}\dots,k_{1},k_{0}}\atop %
         {k_{j}\neq k_{j-1}, j=1,\dots,n}}
   \prod_{i=1}^n\; \int\limits_{\tau_{i-1}}^\tau d\tau_i\;
   \prod_{i=1}^n \int\limits_x^1 dz_i\; 
   \prod_{i=1}^{n+1} \int\limits_x^1 dZ^{(i)}\;
\\ \times&
      e^{-(\tau-\tau_n)R'_k}\;
      {d^A_{k}}^*(\tau,z^{\rm eff}_{\min} | \tau_{n})\;
      \frac{1}{Z^{(n+1)}} d^B_k(\tau, Z^{(n+1)}|\tau_{n})
\\ \times&
      \prod_{i=1}^n \bigg[
      \Peu_{k_ik_{i-1}}^\Theta (z_i)\;
      e^{-(\tau_i-\tau_{i-1})R'_{k_{i-1}}}\;
      {d^{A}_{k_{i-1}}\hskip -3mm}^*\hskip 3mm
      (\tau_i,z^{\rm eff}_{\min} | \tau_{i-1})\;
       \frac{1}{Z^{(i)}} d^B_k(\tau_i, Z^{(i)}|\tau_{i-1})
      \bigg]\;\;
\\ \times&
      \Theta\bigg(\prod_{j=1}^n z_j \prod_{j=1}^{n+1} Z^{(j)} -x\bigg),
\\
  {d^A_{k}}^*(\tau_i,z^{\rm eff}_{\min} | \tau_{i-1}) 
 &= e^{-(\tau_i-\tau_{i-1})R^A_{kk}}
   \sum_{n'^{(i)}=0}^\infty\;
   \prod_{m=1}^{n'^{(i)}}\; 
       \int\limits_{\tau'^{(i)}_{m-1}}^{\tau_i} d\tau'^{(i)}_m\;
   \prod_{m=1}^{n'^{(i)}}\; \int\limits^1_{z^{\rm eff}_{\min}} dz'^{(i)}_m\;
      \bar\Peu_{kk}^{\Theta A} (z'^{(i)}_m)\;
\\ &= \exp\Biggl(
         {(\tau_i-\tau_{i-1})\biggl(-R^A_{kk} 
            +\int\limits^1_{z^{\rm eff}_{\min}} dz\bar\Peu_{kk}^{\Theta A} (z)
                             \biggr)}
          \Biggr),\;\;\; 
          \tau^{(i)}_0=\tau_{i-1},
\\
  d^B_G(\tau_i, Z^{(i)}|\tau_{i-1})
 &= e^{-(\tau_i-\tau_{i-1})R^B_{GG}}
   \bigg\{ \delta(Z^{(i)}-1)+
\\&
  +\sum^\infty_{ n^{\bu(i)}=1}\;\;
     \frac{(\tau_i-\tau_{i-1})^{n^{\bu(i)}}}{n^{\bu(i)}!}
     \prod_{m=1}^{n^{\bu(i)}} 
        \int\limits_0^1 dz_m^{\bu(i)}\;
        \bar\Peu_{GG}^{\Theta B} (z_m^{\bu(i)})\;
     \delta\Big(Z^{(i)} -\prod_{m=1}^{n^{\bu(i)}} z_m^{\bu(i)} \Big)
  \bigg\}.
  \end{split}
\end{equation}
The distribution $d^B_G$ for the trouble-making
component of the kernel 
$\bar P_{GG}^{\Theta B} (z) =P_{GG}^{\Theta B} (z) =2C_A/z$ 
and $R^B_{GG}= (\alpha_S(t_A)/\pi)2C_A$,
can be calculated
analytically:
\begin{equation}
\label{eq:dbg-anal}
\begin{split} 
 d^B_G(\tau_i,Z^{(i)}|\tau_{i-1}) = &
  e^{-(\tau_i-\tau_{i-1}) R^B_{GG}}
  \Bigg[ \delta(Z^{(i)}-1)
\\ &
  +\frac{1}{Z^{(i)}}
   \sum_{n=1}^\infty \frac{(\tau_i-\tau_{i-1})^n}{n!}
    (R^B_{GG})^n\;  \prod_{i=1}^{n} \int_0^1 dz_i\;
  \delta\bigg(Z^{(i)}- \prod_{j=1}^{n} z_j \bigg)
  \Bigg].
\end{split}
\end{equation}
The integral
\begin{equation}
  \prod_{i=1}^{n} \int_0^1 dz_i\; \delta\bigg(Z^{(i)}- \prod_{j=1}^{n} z_j \bigg)
= \prod_{i=1}^{n} \int d\ln z_i\; 
  \delta\bigg(\ln Z^{(i)}- \sum_{j=1}^{n} \ln z_j \bigg)
= \frac{[\ln(1/Z^{(i)})]^{n-1}}{(n-1)!}
\end{equation}
is just the volume of the simplex and when inserting it in 
eq.~(\ref{eq:dbg-anal}) we find
\begin{equation}
 d^B_G(\tau_i,Z^{(i)}|\tau_{i-1}) =
  e^{-(\tau_i-\tau_{i-1}) R^B_{GG}}
  \bigg[ \delta(Z^{(i)}-1)
  +\frac{1}{Z^{(i)}}
    \sum_{n=1}^\infty \frac{(\tau_i-\tau_{i-1})^n}{n!(n-1)!}
    (R^B_{GG})^n\; \ln^{n-1}\frac{1}{Z^{(i)}}
  \bigg].
\end{equation}
The above can easily be expressed in terms of  
the $I_1$ Bessel function: 
\begin{equation}
_0F_1(2;u) =\sum_{n=0}^\infty \frac{u^{n}}{n!(n+1)!}
=\frac{1}{\sqrt{u}}I_1(2\sqrt{u}).
\end{equation}
We shall, however, introduce our own notation:
\begin{equation}
\label{eq:bprim}
B'(\eta,Z^{(i)}) = e^{-\eta}[\delta(1-Z^{(i)})+\eta _0F_1(2;-\eta\ln Z^{(i)})],
\end{equation}
which leads to the simple expression
\begin{equation}
 d^B_G(\tau_i,Z^{(i)}|\tau_{i-1}) =
  \frac{1}{Z^{(i)}}
 B'\Big((\tau_i-\tau_{i-1})R^B_{GG},Z^{(i)}\Big),
\end{equation}
which can easily be plugged into a MC program.


In the above results of the analytical
integrations, we  easily identify the compact analytical expression
for the distributions of the
$z$ and $\tau$ variables of the flavour-changing emissions
(upper layer in the hierarchy). 
For each pure gluonic segment, there is one additional variable $Z$.

Since the average multiplicity of the flavour-changing emissions
is $\sim 1$, we may simply plug in the integrations
over $\tau_i,z_i,Z^{(i)}$, $i=1,...,n_{\max}$ into 
any general-purpose MC simulator, for instance into the {\tt FOAM} program 
\cite{foam:1999, foam:2002}. 
The value of $n_{\max}=5$ is probably more than sufficient
for a precision of $10^{-4}$ and it is feasible for {\tt FOAM} 
(up to about 20-dimensional distributions),
especially because the integrand does not involve any strong singularities.
Also, generating points according to
the higher-dimensional distributions will be done very rarely.


This completes the theoretical description of the constrained MC 
algorithm of type II.b.

\subsection{Construction of non-Markovian constrained MCs, type II}
In this section we present an actual implementation of some of 
the constrained MC algorithms of class II described in the previous 
sections. Some numerical results are also given.

We shall proceed from simple examples of the MC algorithms
for simplified distributions,
gradually going to more elaborate examples in which 
the previous, simpler, MC examples are used as benchmarks 
in the numerical tests.
It is worthwhile to describe the above step-by-step method
of creating more and more sophisticated versions of the MC algorithm and its
numerical realization, because 
it is an essential part of constructing any {\em precision MC event generator},
albeit it is rarely explicitly exposed in the literature. It
can be of vital interest for any reader interested in the practical aspects
of constructing MC event generators%
\footnote{Such simplified MC programs existed for many precision MCs
  for the QED calculations with YFS exclusive exponentiation;
  see for instance ref.~\cite{Jadach-yfs-mpi:1987}.}.

\subsubsection{Benchmark MC for $P_{GG}=2C_A/z$, Poisson-type and inefficient}
\begin{figure}[!ht]
  \centering
  {\epsfig{file=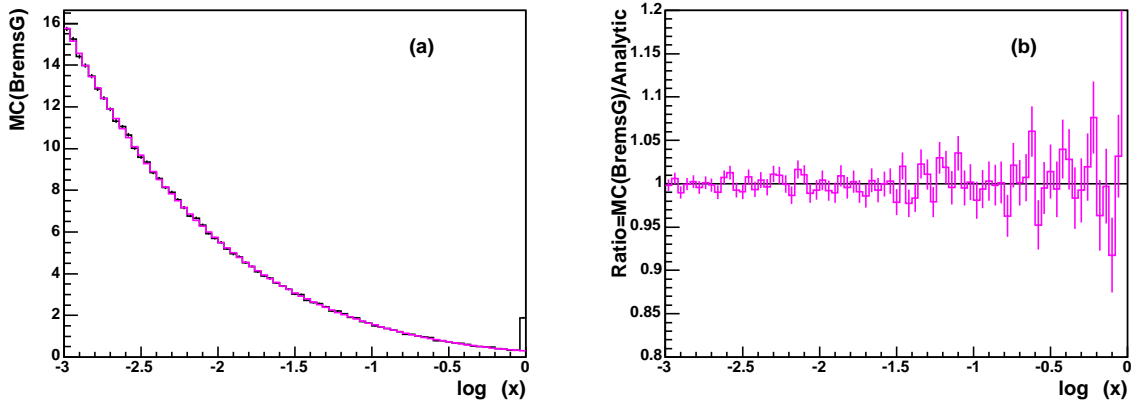,width=160mm}}
  \caption{\sf
    Pure gluon case. $P_{GG}=2C_A/z$.
    The distribution of $x=\prod_i z_i$ and its ratio to 
    the analytical prediction $B'(\gamma,x)$.
    }
  \label{fig:besselMC}
\end{figure}

As a warming-up exercise, let us now work out in detail a MC
algorithm calculating the following integral,
cf.\ eq.~(\ref{eq:dbg-anal}):
\begin{equation}
  I(\gamma)= 
  \int_{\epsilon_1}^1 dx \; d^B_G(\tau,x|\tau_0)
 =\int_{\epsilon_1}^1 dx \frac{1}{x}\; B'(\gamma,x),
\end{equation}
where $\gamma=(\alpha_S(t_A)/\pi)2C_A (\tau-\tau_0)$ and $\epsilon_1\ll 1$.
\begin{equation}
 B'(\gamma,x) =
  e^{-\gamma}
  \Bigg[ \delta(x-1)
  +\sum_{n=1}^\infty \frac{\gamma^n}{n!}
   \prod_{i=1}^{n} \int_0^1 dz_i\;
  \delta\bigg(x- \prod_{j=1}^{n} z_j \bigg)
  \Bigg]
\end{equation}
with the aim of preparing basic tools and setting baseline normalization
for the MC algorithm of type II.b
(similarly as it was done in ref.~\cite{Jadach-yfs-mpi:1987}).
On the one hand, the Bessel-class function $B'$ is known analytically 
in terms of a series (\ref{eq:bprim}).
On the other hand, the integral $I(\gamma)$ can be rewritten as
\begin{equation}
 I(\gamma) = 
  e^{-\gamma}
  \Bigg[ 1
  +\sum_{n=1}^\infty \frac{\gamma^n}{n!}
   \prod_{i=1}^{n} \int_{\epsilon_1}^1 \frac{dz_i}{z_i}\;
  \Theta\bigg(\prod_{j=1}^{n} z_j -\epsilon_1 \bigg)
  \Bigg],
\end{equation}
remembering that $x=\prod_i z_i$.
The above integral is easily implementable in the MC, which treats the $\Theta$
function as a MC weight: $W=\Theta\Big(\prod_{j=1}^{n} z_j -\epsilon_1 \Big)$.
The variable $n$ is generated according to the distribution
\begin{equation}
  \bar I_0 =e^{-\gamma},\;
  \bar I_{n>0}= \frac{e^{-\gamma} \gamma^n}{n!} \ln^n \frac{1}{\epsilon_1},~~~~
  \bar I=\sum_{n=0}^\infty \bar I_n =e^{-\gamma}e^{-\gamma\ln\epsilon_1}.
\end{equation}
The variables $z_i\in (\epsilon_1,1)$ are generated according to
the distribution $1/z_i$.
Once we generate suitably long series of MC events $(n;z_1,z_2,...,z_n)$
we calculate the integral using the average weight, with the usual expression
$I=  \langle W \rangle  \bar I$.
In the same MC run we can also obtain the distribution $B'(\gamma,x)/x$,
just by examining the histogram of $x=\prod_i z_i$.

In the LHS plot of fig.~\ref{fig:besselMC} we show the (properly normalized)
distribution of $x$ from the MC.
The acceptance rate $\sim 3\times 10^{-4}$ 
is rather low  -- it demonstrates the problem
with the $1/z$ component in any MC (also Markovian) in which
the starting point of the generation of the emission probability is of the 
Poisson type. 
This phenomenon is quite general.
Our numerical example shows the evolution from $Q=1$ GeV to $Q=1$ TeV.
The average emission multiplicity in the MC run is about 3.4 
for $\epsilon_1=10^{-3}$.
Since the resulting distribution of $x$
is known analytically, we can also examine its ratio to the MC result.
In the RHS plot of fig.~\ref{fig:besselMC} we show this ratio 
(for $1.4\times 10^{9}$ events). 
It is equal to 1, to within the statistical error of order $\sim 1\%$.
In the LHS plot we clearly see that the contribution $\sim \delta(x)$
is reproduced by this MC algorithm/program
(absent in the analytical program).

\begin{figure}[!ht]
  \centering
  {\epsfig{file=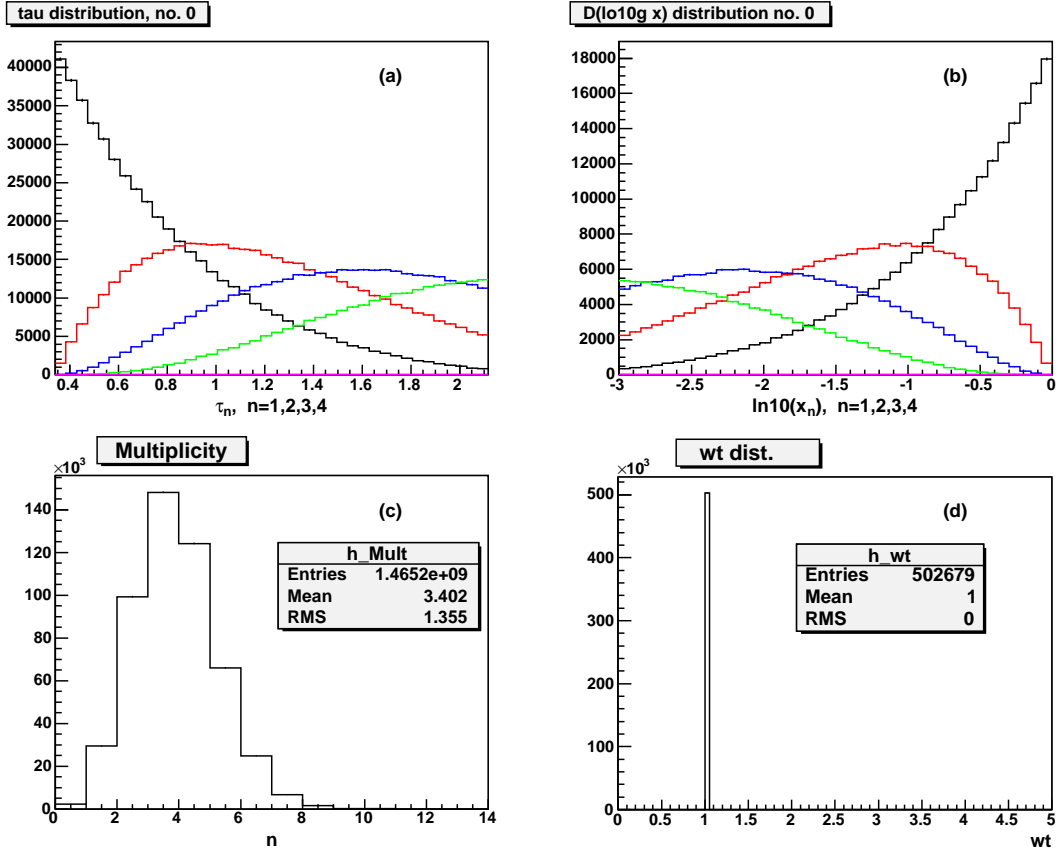,width=150mm}}
  \caption{\sf
    Pure gluon case. Kernel$=2C_A/z$.
    Distribution of the variables $x_i=\prod_{n=1}^i z_i$
    and of the ordered $\tau_i\in(\tau_0,\tau)$.
    }
  \label{fig:besselMC2}
\end{figure}
For the purpose of the next exercises
we are interested not only in the value of the integral, but also in
the exclusive distributions.
In fig.~\ref{fig:besselMC2} we examine the distributions of the first four
variables $x_i=\prod_{n=1}^i$ and $\tau_i$ in the emission chain.
The ordered $\tau_i$ variables are generated within the range $(\tau_0,\tau)$
corresponding to $Q_0=1$ GeV and $Q=1$ TeV.
In the following we shall check that the above semi-exclusive distributions
are correctly reproduced by more sophisticated MC algorithms.
In this figure we also include the distributions of the emission
multiplicity and the MC weight. 
In the weight distribution we exclude zero-weight events.

\begin{figure}[!ht]
  \centering
  {\epsfig{file=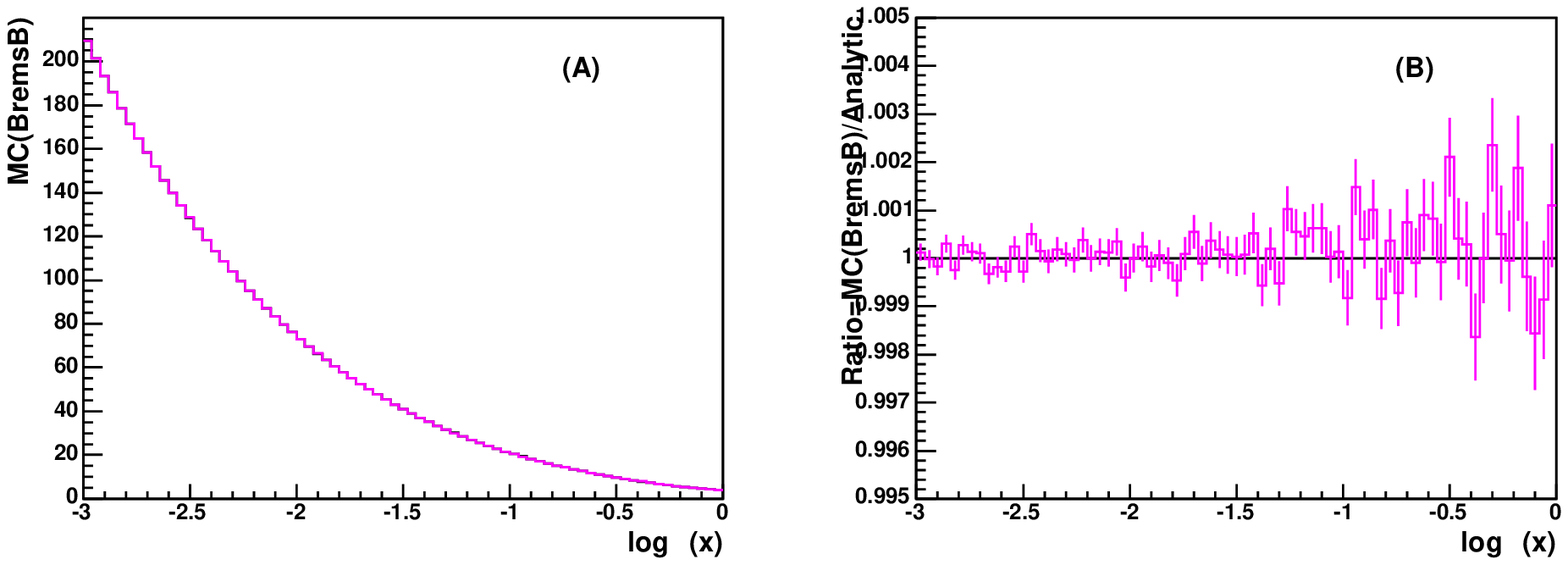,width=150mm}}\\
  {\epsfig{file=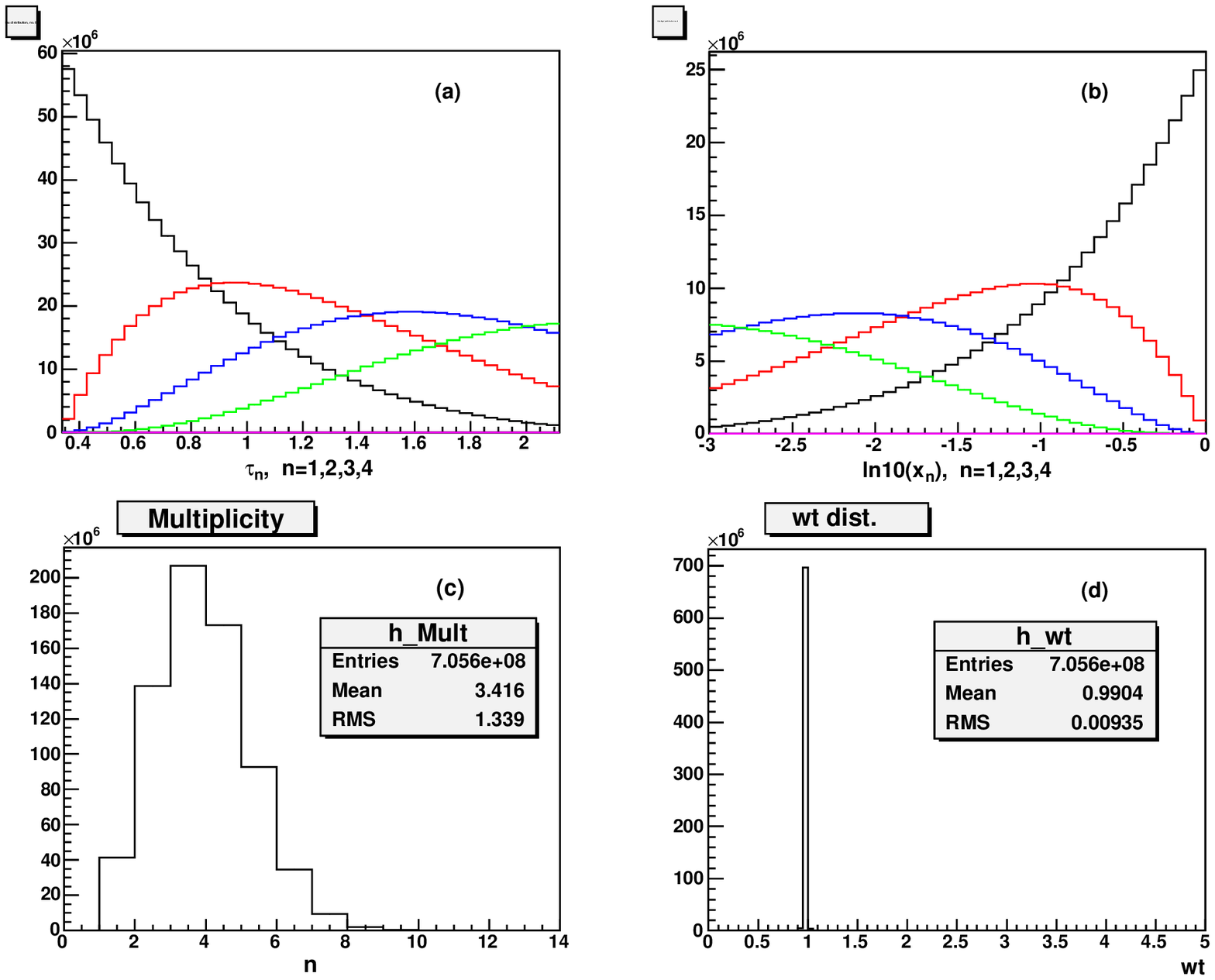,width=150mm}}
  \caption{\sf
    Pure gluon case. Kernel$=2C_A/z$.
    Distribution of the variables $x_i=\prod_{n=1}^i z_i$
    and of the ordered $\tau_i\in(\tau_0,\tau)$. 
    }
  \label{fig:besselMC4}
\end{figure}

\subsubsection{Weight-1 algorithm for $P_{GG}=2C_A/z$, Bessel type}
The inefficiency of the algorithm described in the
previous subsection is mainly due to the fact that
the emission probability distribution in the integral under consideration
is of the type
$P_n\sim x^n/n!(n-1)!$, Bessel-type for short, while in the MC we actually
generate a Poisson distribution $P_n\sim x^n/n!$ and turn it into 
a Bessel-type one
by the inefficient brute-force rejection method. 
Now we proceed to the next step -- we construct
a prototype algorithm in 
which a Bessel-type emission probability
is used from the start and there is no need for the rejection at all.
The previous inefficient Poisson-type MC will be useful, however,
as a precision cross-check for the new one, 
especially for testing semi-exclusive distributions.

Let us consider almost the same integral
\begin{equation}
  I'(\gamma)= 
  \int_{\epsilon_1}^1 dx \frac{1}{x}\; h(x)\; 
  \gamma \;\; _0F_1(2;-\gamma\ln(x)),
\label{eq:Iprim}
\end{equation}
which in the multi-integral form looks as follows:
\begin{equation}
 I'(\gamma) = \int_{\epsilon_1}^1 dx \frac{1}{x}\; h(x)\;
  \Bigg[
  \sum_{n=1}^\infty \frac{\gamma^n}{n!}
   \prod_{i=1}^{n} \int_0^1 dz_i\;
  \delta\Bigg(x- \prod_{j=1}^{n} z_j \Bigg)
  \Bigg],
\end{equation}
where we have removed the unimportant $\delta(1-z)$ component and inserted
the test function%
\footnote{In the following numerical exercises we set it
  to the constant value $h(x)=1.908359$.}
$h(x)/x$.
This integral can be rewritten as 
\begin{equation}
  I'(\gamma) = \int\limits_{\epsilon_1}^1 \frac{dx}{x}
  h(x)
  \bigg[
    \sum_{n=1}^\infty
    \frac{\gamma^n \ln^{n-1}(1/x) }{n!(n-1)!}
    \biggl(
      \frac{(n-1)!}{\ln^{n-1}(1/x)}\prod_{i=1}^{n} 
      \int\limits_0^1
      d\ln z_i\;    
      \delta\Big(\ln x- \sum_{j=1}^{n} \ln z_j \Big)
    \biggr)\bigg],
\end{equation}
where the internal part of the integrand is conveniently normalized as
\begin{equation}
    \frac{(n-1)!}{\ln^{n-1}(1/x) }\;    
    \prod_{i=1}^{n} \int_0^1 d\ln z_i\;
    \delta\Big(\ln x- \sum_{j=1}^{n} \ln z_j \Big)
    \equiv 1
\end{equation}
and we may simulate $z$ variables very easily.
Changing variables to $y_i=\ln z_i$, we see that
the distribution in $y_i$ is a uniform distribution over the
$n-1$ dimensional simplex.
There are several convenient methods of generating points uniformly
within such a simplex.
The simplest method is to throw 
randomly $n-1$ uniform points $u_i\in(\ln x, 0),$  $i=1,2,\dots,n-1$ 
and order them using any standard method: 
$\ln x=u_n< u_{n-1} <\dots <u_1<u_0=0$. 
Then we take the differences $y_i=u_{i}-u_{i-1}$, $i=1,\dots,n$,
which by construction fulfil the constraint  $\ln x=\sum_{j=1}^{n} y_j$.

The MC algorithm consists of the following steps:
first the $x$ is generated according to the distribution
\begin{equation}
\rho(x) =\frac{1}{x}\; h(x)\; 
  \gamma\;\; _0F_1(2;-\gamma\ln(x)).
\end{equation}
Next the number of emissions $n$ is generated according to a
(normalized) Bessel-type probability%
\footnote{It is done by using the simple/universal method of inverting
  cumulative distribution.}
distribution%
\footnote{
Let us note that a similar Bessel-type distribution of 
the number of emissions is used by Kharraziha and Lonnblad 
in the event generator based on 
the Linked Dipole Chain model \cite{Kharraziha:1997dn}.}:
\begin{equation}
P_n =
  \Bigl(\gamma\;\; _0F_1(2;-\gamma\ln(x))\Bigr)^{-1}
    \frac{\gamma^n \ln^{n-1}(1/x) }{n!(n-1)!},~~~ \sum_{n=1}^\infty P_n=1.
\end{equation}
Finally the variables $z_i$ are generated as described above.
In this algorithm all events are generated with weight 1,
provided $x$ is generated exactly according to $\rho(x)$, for instance
using the general-purpose tool {\tt FOAM}.
The algorithm is very efficient and fast.

Numerical results from the corresponding MC program 
are shown in fig.~\ref{fig:besselMC4}. 
Plot (A) in this figure shows the distribution which is
the integrand of eq.~(\ref{eq:Iprim}) from the MC run.
The analytical result superimposed on the same plot is indistinguishable from
the MC result.
In the next plot (B) we show the ratio of the two distributions,
MC and analytical. 
They agree within a very small statistical error of order $\sim 10^{-4}$.
In the next two plots, (a) and (b), we see that the new algorithm
reproduces perfectly well the semi-exclusive distributions of
the same two plots in fig.~\ref{fig:besselMC2}.
The multiplicity distribution in the plot (c) is also well reproduced.
Plot (d) shows the MC weight distribution.

\begin{figure}[!ht]
  \centering
  {\epsfig{file=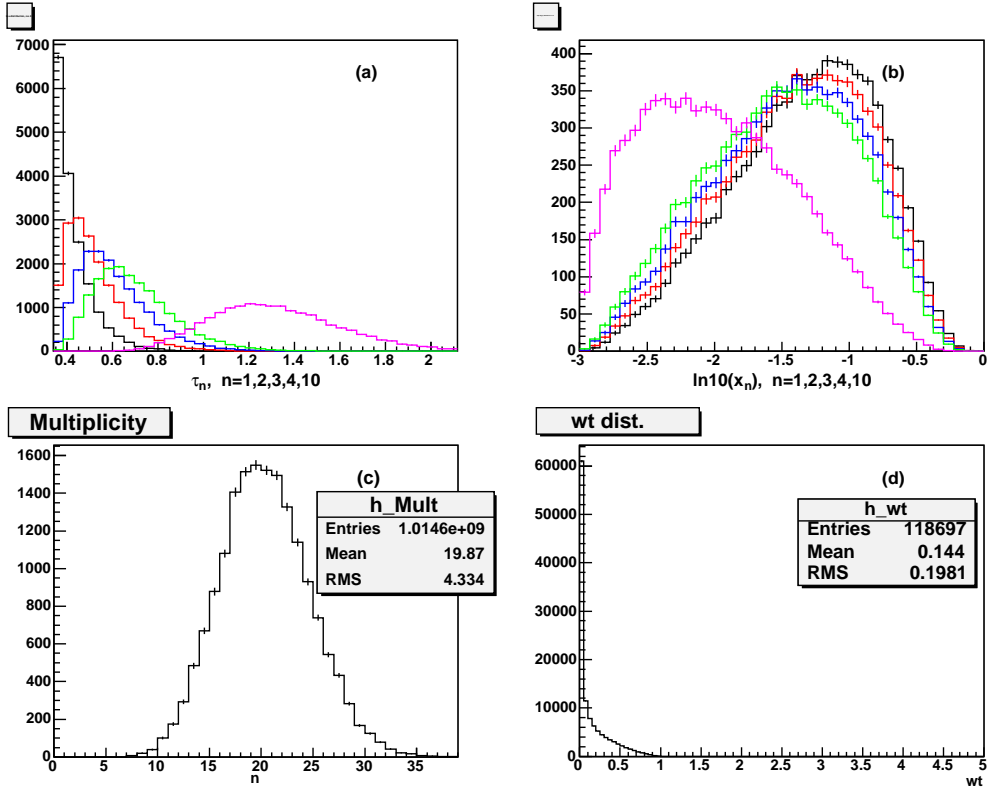,width=140mm}}
  \caption{\sf
    Results from {\tt BremsP} for the gluon emitter, $k=G$.
    }
  \label{fig:BremsPggProt}
\end{figure}
\begin{figure}[!ht]
  \centering
  {\epsfig{file=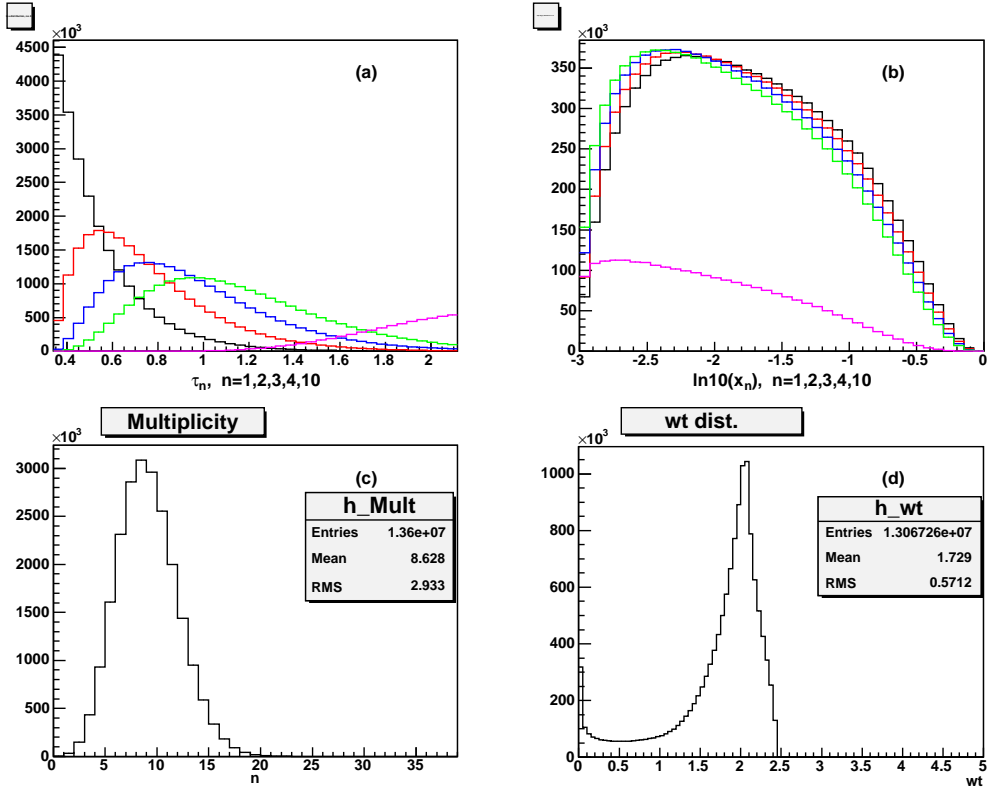,width=140mm}}
  \caption{\sf
    Results from {\tt BremsP} for quark emitter, $k=Q$.
    }
  \label{fig:BremsPqqProt}
\end{figure}
\begin{figure}[!ht]
  \centering
  {\epsfig{file=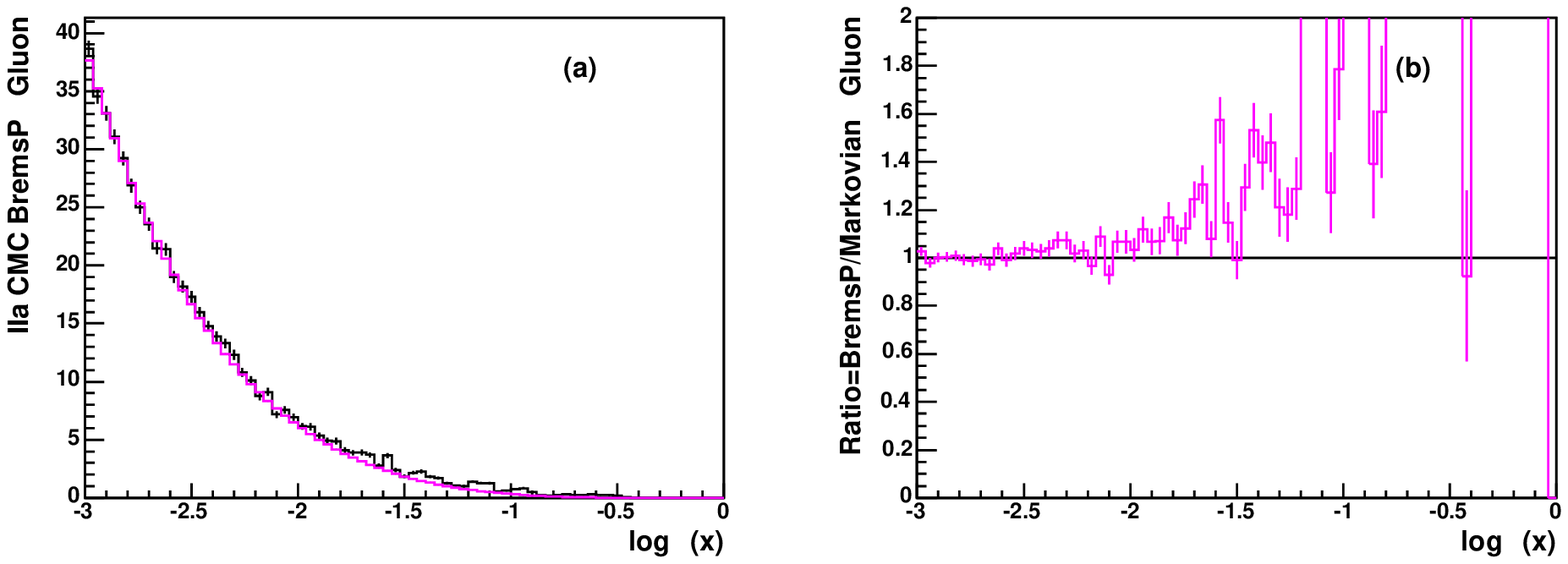,width=140mm}}\\
  {\epsfig{file=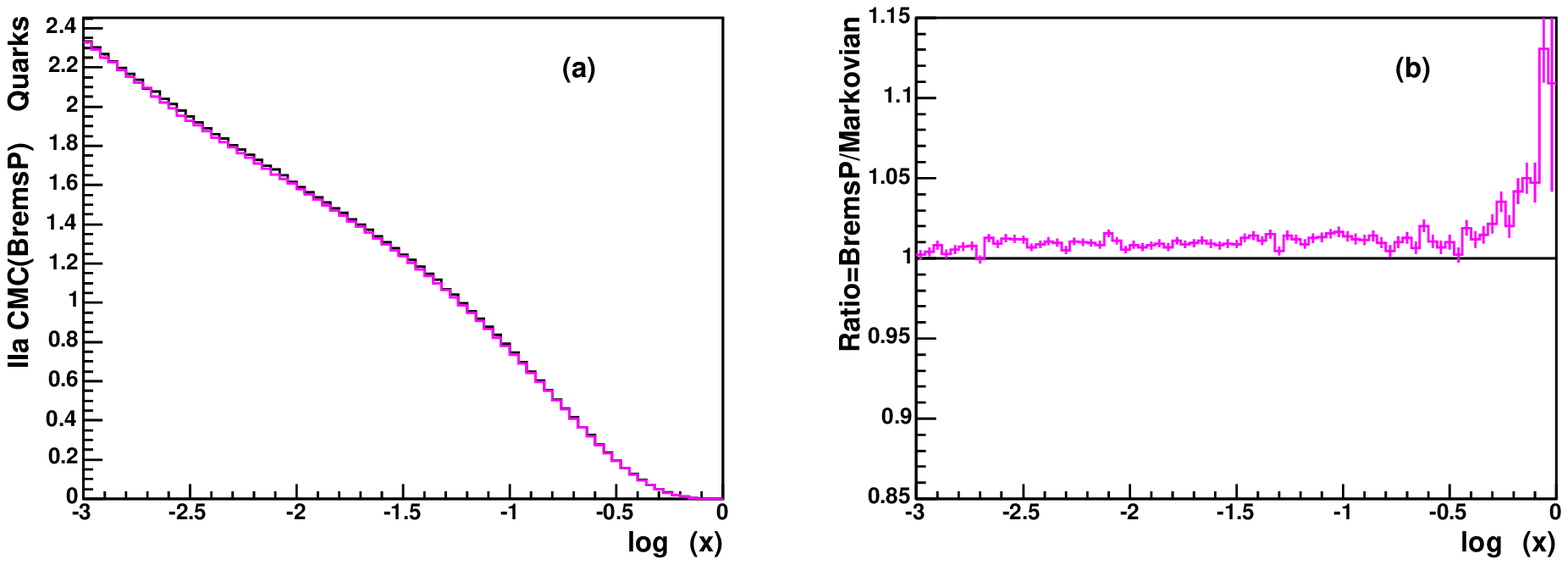,width=140mm}}
  \caption{\sf
    Results from the non-Markovian {\tt BremsP} (II.a)
    compared with results of the Markovian {\tt EvolMC}.
    Evolution from 1 GeV to 1 TeV due to multiple gluon emission
    from the gluon emitter line (upper plots) 
    and quark emitter line (lower plots).
    Starting $D_k(\tau_0,z)$ as in the realistic proton.
    $N_f=3$ in the virtual form factor.
    } 
  \label{fig:Mark_step1}
\end{figure}
\subsubsection{Prototype benchmark type II.a, pure bremsstrahlung}
The two toy MCs from previous sections should be regarded as 
introductory exercises (and numerical benchmarks)
for the next step,
in which we shall elaborate on the constrained Markovian MC solution
with $x$-tagging of the type II.a.
We shall restrict ourselves to pure brems\-strahlung from the
gluon or quark line, 
without using the multibranching to isolate $P^B_{GG}\sim 1/z$.
The corresponding MC prototype we name {\tt BremsP}.
The purpose of that is threefold:
(a) to measure the MC efficiency of this class of the MC algorithms, 
(b) to provide a cross-check for the 
more sophisticated prototype MC with multibranching for the 
brems\-strahlung from the gluon line, which will be developed in the next
section,
(c) to compare it with the other constrained Markovian MC prototypes
for the pure brems\-strahlung.

The starting point for the construction of the algorithm is 
eq.~(\ref{eq:sc5}). 
Its simplified version,
restricted to the pure bremsstrahlung case, is the following:
\begin{equation}
  \label{eq:bremsP1}
  \begin{split}
  &\sigma_k
  =\int dx\; 
   H_k(x)\; A_{k} x^{-1+\eta}\;
   e^{-(\tau-\tau_0)R_{kk}}
  \sum_{n=0}^\infty \;
   \prod_{i=1}^n\;\; \int\limits_{\tau_{i-1}}^\tau d\tau_i\;
                      \int\limits_x^1 dz_i\;
      \Peu_{kk}^\Theta (z_i)\;\;
\\&~~~~~~~~~~~~~~~~~~~~~~~~\times
      W^{\Theta}\;
      W^{D}_k(x,\bar x_0),
  \end{split}
\end{equation}
where
\begin{equation}
   W^{D}_{k}(x,x_0) 
   = \frac{D_{k}(\tau_0,x_0)}{A_{k} x_0^{-1+\eta}} 
   \bigg\{ \frac{x}{\bar{x}_0} \bigg\}^{-\eta},~~~~~
   W^{\Theta}= \Theta \Bigg(\prod_{j=1}^n z_j -x\Bigg).
\end{equation}
Neglecting $W^{\Theta}W^{D}$
we can perform a $z$-integration and $n$-summations:
\begin{equation}
\sigma_k
  =\int_0^1 dx\; 
   H_k(x)\; A_{k} x^{-1+\eta}\; 
   e^{(\tau-\tau_0)(-R_{kk}+\Omega_{kk}(x))},~~~~
   \Omega_{kk}(x)
      =\int\limits_x^1 dz\; \Peu_{kk}^\Theta (z).
\end{equation}
The emission multiplicity distribution is Poissonian:
\begin{equation}
  P_n(x)=\frac{1}{n!} e^{-\lambda(x)} \lambda(x)^n,~~~~~
  \lambda(x) =  (\tau-\tau_0) \Omega_{kk}(z)
\end{equation}
and we may generate it together
with the $\tau_i$ variables, much as in the Markovian case,
except that the average multiplicity (forward leap in Markovian random walk)
now depends on $x$ (in the unconstrained Markovian it was constant).
The variable $x$ is generated as a first variable using {\tt FOAM}
then $n$ and finally $z_i\in (x,1)$ {\em exactly }
according to $\Peu_{kk}^\Theta (z)$.

A few comments on the form factor are in order here.
The part $(\tau-\tau_0) \Omega_{kk}(x)$ is clearly
coming from the real emission and, for instance,  
will be different if we generate according to an approximate 
$\bar \Peu_{kk}^\Theta (z)$;
see later in this section.
The part $-(\tau-\tau_0)R_{kk}=-\Phi_{kk}(\tau,\tau_0)$
is a genuine virtual part of the form factor,
independent of any details of the MC generation, cf.\ eqs.\ 
(\ref{eq:1formfaktor})--(\ref{eq:formfaktory}).
With the usual expansion
\begin{equation}
\label{eq:Rexpansion}
  P_{ik}(\tau,z)= 
     \delta(1-z)      \delta_{ik} A_{kk}
    +\frac{1}{(1-z)_+}\delta_{ik} B_{kk}
    +\frac{1}{z}                  C_{ik}
                                 +D_{ik}(z),
\end{equation}
we obtain
\begin{equation}
\label{eq:Rkkexpansion}
  R_{kk}=(\tau-\tau_0)^{-1}\Phi_{kk}(\tau,\tau_0)
   =\frac{\alpha_S(t_A)}{\pi} \bigg[
     B_{kk} \ln\frac{1}{\epsilon}
    -A_{kk}    \bigg],
\end{equation}
and the real emission form factor
is
\begin{equation}
  \Omega_{kk}(x)=
   \frac{\alpha_S(t_A)}{\pi} \bigg[
     B_{kk} \ln\frac{1-x}{\epsilon}
    +C_{kk} \ln\frac{1}{x}
    +\int_x^1 dz\; D_{kk}(z) 
    \bigg],
\end{equation}
where
\begin{equation}
 \int_x^1 dz\; D_{GG}(z)
  =2C_A \bigg(-\frac{11}{6} +x\Big( 2 -\frac{1}{2}x +\frac{1}{3}x^2\Big)\bigg)
\end{equation}
and
\begin{equation}
 \int_x^1 dz\; D_{qq}(z)
  =C_F \bigg(-\frac{3}{2}+x+\frac{1}{2}x^2 \bigg).
\end{equation}

In fig.~\ref{fig:BremsPggProt}, we show type II.a MC results 
for the same semi-exclusive distributions as previously,
using realistic gluon distribution $D_G(\tau_0,x)=C x^{-0.8}(1-x)^5$,
for the gluon-strahlung out of the gluon emitter line.
As we see, the efficiency of the MC is extremely low --
the acceptance rate is merely  $1.5\times 10^{-5}$ 
(note that the weight-0 events are not included in 
fig.~\ref{fig:BremsPggProt}d).
Nevertheless, these results 
will still be useful to cross-check the more efficient algorithm type II.b
in the next section.
We have investigated what the sources of the inefficiency are.
As in the previous toy model, the main reason for low efficiency 
is that there are many zero-weight events due to $W^{\Theta}$.
The factor $(1-x)^5$ in the gluon distribution
causes a loss of efficiency of a factor 3.
The factor $x^{-0.8}$ accounts for a mere factor 2 in the efficiency loss.
It is therefore not urgent to eliminate this efficiency by means of
incorporating $z^{-\eta}$ factor into $D_{GG}(z)$.
This possibility we have considered in the general discussion on method II.
On the other hand, a factor 3 loss in the MC efficiency in method II, which is
due to the presence of $(1-z)^5$ in the gluon SF, 
looks at first sight {\em irreducible}.
Nonetheless, one may consider modelling this factor
using the internal rejection loop, because 
the $(1-z)^5$ factor, upon expanding, is a sum of monomials $z^p$ and 
the overall normalization can be calculated 
(with non-MC methods) as a sum over these terms.
It is not excluded that
with some extra effort, the overall efficiency of the method II.a
could be improved to the level of $10^{-4}$.

Let us now repeat the same exercise for the brems\-strahlung
emitted from the quark line. 
In fig.~\ref{fig:BremsPqqProt} we show the corresponding
results ($k=q$) and the starting quark distribution
being $D_q(\tau_0,x)=D_{sea}(z)+D_U(z)+D_D(z)$, that is sea plus both
valence quarks, taking a typical parametrization of the proton 
parton distribution function at $Q_0=1$ GeV.
Strikingly, the overall efficiency is very good;
the rejection rate $\simeq \langle w\rangle /\langle w_{\max}\rangle $ 
is only about $30\%$!
Obviously, without $1/z$ component in the kernel,
the basic algorithm type II is quite efficient.
It should be remembered that in the actual run
$P_{qq}(z)$ is generated exactly 
(i.e.\ with the help of the internal rejection loop).
The fact that the weight distribution extends above 1,
up to 2.5, is related to the valence component.
However, the entire weight distribution looks very well
for the optional rejection method.

The overall normalization of this MC is cross-checked with the help
of the Markovian MC {\tt EvolMC} of ref.~\cite{Jadach:2003bu}.
In the top part of fig.~\ref{fig:Mark_step1} we show result of the evolution
from 1 GeV to 1 TeV in which we restrict ourselves to gluon emission
out of the gluon line, taking the starting gluon distribution as in the proton.
The non-Markovian type II.a MC {\tt BremsP}
reproduces the results of the Markovian MC {\tt EvolMC}
within a statistical error of a few per cent.
The apparent discrepancy at high $x$ values is most likely due to some
technical bias related to extremely high MC event rejection rate%
\footnote{We did not try to investigate its precise source, because
  the practical importance of {\tt BremsP} is limited to a test
  of semi-exclusive distributions, not normalization.}.

In the low part of fig.~\ref{fig:Mark_step1} we present the analogous
comparison of {\tt BremsP} and {\tt EvolMC} for multiple 
gluon emission from the quark line.
Again, the agreement is quite reasonable, this time within a smaller
statistical error of $\sim 1\%$.

As an additional cross-check
we also implemented another variant of the II.a type 
constrained MC algorithm {\tt BremsP},
with the approximate kernels $\hat \Peu$
and correcting weight applied at the very end of the MC generation.
In this case we define
\begin{equation}
  \label{eq:bremsP2}
  \begin{split}
  &\sigma_k
  =\int dx\; 
   H_k(x)\; A_{k} x^{-1+\eta}\;
   e^{-(\tau-\tau_0)R_{kk}}
  \sum_{n=0}^\infty \;
   \prod_{i=1}^n\;\; \int\limits_{\tau_{i-1}}^\tau d\tau_i\;
                      \int\limits_x^1 dz_i\;
      \hat\Peu_{kk}^\Theta (z_i)\;\;
\\&~~~~~~~~~~~~~~~~~~~~~~~~\times
      W^{\Theta}\;
      W^{D}_k(x,\bar x_0)\;
      W^{\Peu}_{k},
  \end{split}
\end{equation}
where the additional weight is
\begin{equation}
   W^{\Peu}_{k}
   = \prod_{i=0}^n 
     \frac{P^\Theta(z_i)}{\hat P^\Theta(z_i)}.
\end{equation}
Neglecting weights we have
\begin{equation}
\sigma_k
  =\int_0^1 dx\; 
   H_k(x)\; A_{k} x^{-1+\eta}\; 
   e^{(\tau-\tau_0)(-R_{kk}+\hat\Omega_{kk}(x))},~~~~
   \hat\Omega_{kk}(x)
      =\int\limits_x^1 dz\; \hat\Peu_{kk}^\Theta (z),
\end{equation}
where the simplified kernel is defined as 
\begin{equation}
  \hat P_{kk}(t,z)= 
     \delta(1-z)       A_{kk}
    +\frac{1}{(1-z)_+} B_{kk}
    +\frac{1}{z}       C_{kk},
\end{equation}
leading to the following real emission form factor
\begin{equation}
  \hat\Omega_{kk}(x)=
   \frac{\alpha_S(t_A)}{\pi} \bigg[
     B_{kk} \ln\frac{1-x}{\epsilon}
    +C_{kk} \ln\frac{1}{x}
    \bigg].
\end{equation}
We have checked that the above MC algorithm gives the same quark
and gluon distributions, as expected.
It is also quite interesting to check how strongly the efficiency of the MC
deteriorates when the additional weight $W^{\Peu}$ is introduced.
In the quark case, the acceptance rate drops from 0.7 to 0.25,
which is not much,
while for gluons it drops  by a factor $\sim 10$, well below $10^{-5}$.

In the next step we will clone the MC subgenerator of type II,
which generates brems\-strahlung from the quark line 
according to simplified $\bar P^\Theta_{qq}=2C_F/(1-z)$
and from the gluon line according to ``truncated'' simplified  
$\bar P^{\Theta A}_{GG}=2C_A/(1-z)$.
After that, having tested the components at hand,
we shall introduce the integration over $Z$ using {\tt FOAM}
and for the brems\-strahlung from the gluon line
we shall combine the Bessel's MC with $P^{\Theta B}_{GG}=2C_A/z$
with the above MC for $P^{\Theta A}_{GG}=2C_A/(1-z)$
and compare resulting distributions 
with the Markovian benchmark of fig.~\ref{fig:Mark_step1}.
This will close the most important first step in making a prototype 
MC according to method II.b.

\subsubsection{Constrained MC type II.b for pure bremsstrahlung}
In the following we implement the algorithm II.b in the case of pure
bremsstrahlung from the gluon or quark line.
In this particular case, the master formula of eq.~(\ref{eq:II.bmc}) 
for the early stage MC 
(obtained from eq.~(\ref{eq:IIBmaster}) by neglecting the MC weight)
has only one variable $Z^{(1)}$ and $z^{\rm eff}_{\min}=x/Z^{(1)}$.
It takes the following simplified form:
\begin{equation}
  \begin{split}
  \bar\sigma_k
  =\int_0^1 dx\; &  H_k(x)\; 
   A_{k} x^{-1+\eta}\;
   \int\limits_x^1 dZ^{(1)}\;
      e^{-(\tau-\tau_0)R'_k}\;
      {d^{A}_{k}}^*(\tau,z^{\rm eff}_{\min} | \tau_{0})\;
      \frac{1}{Z^{(1)}} d^B_k(\tau, Z^{(1)}|\tau_{0}),
\\
  {d^{A}_{k}}^*(\tau,z^{\rm eff}_{\min} | \tau_{0}) 
 &= e^{-(\tau-\tau_{0})R^A_{kk}}
   \sum_{n'=0}^\infty\;
   \prod_{m=1}^{n'}\; 
       \int\limits_{\tau'_{m-1}}^{\tau} d\tau'_m\;
   \prod_{m=1}^{n'}\; \int\limits^1_{z^{\rm eff}_{\min}} dz'_m\;
      \bar\Peu_{kk}^{\Theta A} (z'_m)\;
\\ &= \exp\Biggl(-(\tau-\tau_{0})R^A_{kk} 
         +(\tau-\tau_{0})\int\limits^1_{z^{\rm eff}_{\min}} dz
                                \bar\Peu_{kk}^{\Theta A} (z)\Biggr)
\\
  d^B_q(\tau, Z^{(1)}|\tau_{0})
 &= \delta(1-Z^{(1)})
\\
  d^B_G(\tau, Z^{(1)}|\tau_{0})
 &= e^{-(\tau-\tau_{0})R^B_{GG}}
   \bigg\{ \delta(Z^{(1)}-1)+
\\&
  +\sum^\infty_{n^{\bu}=1}\;\;
     \prod_{m=1}^{n^{\bu}}
        \int\limits^{\tau}_{\tau^{\bu}_{m-1}} d\tau^{\bu}_m \;
     \prod_{m=1}^{n^{\bu}} 
        \int\limits^1_0 dz_m^{\bu}\;
        \bar\Peu_{GG}^{\Theta B} (z_m^{\bu})\;
     \delta\Big(Z^{(1)} -\prod_{m=0}^{n^{\bu}} z_m^{\bu} \Big)
  \bigg\}
\\& 
  =\frac{1}{Z^{(1)}}
   B'\Big((\tau-\tau_{0})R^B_{GG},Z^{(1)}\Big)
\\& 
  = e^{-(\tau-\tau_{0})R^B_{GG}}
   \bigl(
    \delta(Z^{(1)}-1)
   +\hat d^B_G(\tau, Z^{(1)}|\tau_{0})
   \bigr),
\\
   \hat d^B_G(\tau, Z^{(1)}|\tau_{0})
  &=
   \frac{1}{Z^{(1)}}
    (\tau-\tau_{0})R^B_{GG}\; 
         _0F_1\bigl(2;-(\tau-\tau_{0})R^B_{GG}\ln(Z^{(1)})\bigr).
\end{split}
\end{equation}

The integral proportional to $\delta(1-Z^{(1)})$
has to be treated separately%
\footnote{The need of treating the $\delta$-part separately
  will be more annoying in the general case, with several
  gluon emitter bremsstrahlung segments, because this causes
  proliferation of the separate MC branches with different
  distributions, adding a lot of code, difficult to write and debug.}:
\begin{equation}
  \begin{split}
  \bar\sigma_q
  &=\int_0^1 dx\;   H_q(x)\; 
   A_{q} x^{-1+\eta}\;
      e^{-(\tau-\tau_0)R'_q}\;
      {d^A_{q}}^*(\tau,x| \tau_{0}),
\\
  \bar\sigma_G
  &=\int_0^1 dx\;   H_G(x)\; 
   A_{G} x^{-1+\eta}
   \int\limits^1_x \frac{dZ^{(1)}}{Z^{(1)}}
      e^{-(\tau-\tau_0)R'_G}\;
      {d^A_{G}}^*\Bigl(\tau,\frac{x}{Z^{(1)}} \Big| \tau_{0}\Bigr)\;
      e^{-(\tau-\tau_0)R^B_{GG}}
      \hat d^B_G(\tau, Z^{(1)}|\tau_{0})
\\
  &+\int\limits_0^1 dx\;   H_G(x)\; 
   A_{G} x^{-1+\eta}\;
      e^{-(\tau-\tau_0)R'_G}\;
      {d^A_{G}}^*(\tau,x | \tau_{0})
      e^{-(\tau-\tau_0)R^B_{GG}}.
\end{split}
\end{equation}
The distribution of the variables $x$ and $Z=Z^{(1)}$ for 
the general-purpose simulator {\tt FOAM} are given 
by the integrands in the integrals:
\begin{equation}
\label{eq:sigmabar}
  \begin{split}
  \bar\sigma_q
  &=\int\limits_0^1 dx 
    \int\limits^1_x dZ\;
  H_q(x)\; 
   A_{q} x^{-1+\eta}\;
      \exp\Biggl[-(\tau-\tau_0)
          \biggl(R'_q
         +R^A_{qq} 
         -\int\limits^1_{x} dz
                                \bar\Peu_{qq}^{\Theta A} (z)
          \biggr)\Biggr],
\\
  \bar\sigma_G
  &=\int\limits_0^1 dx 
    \int\limits^1_x \frac{dZ}{Z}\;
  H_G(x)\; 
   A_{G} x^{-1+\eta}\;
\\& \hskip 3cm
      \exp\Biggl[-(\tau-\tau_0)\biggl(R'_G
         +R^A_{GG} 
         +R^B_{GG} 
         -\int\limits^1_{x/Z} dz
                               \bar \Peu_{GG}^{\Theta A} (z)
          \biggr)\Biggr]
       \hat d^B_G(\tau, Z|\tau_{0})
\\ &
   +\int\limits_0^1 dx\;   H_G(x)\; 
   A_{G} x^{-1+\eta}\;
      \exp\Biggl[-(\tau-\tau_0)\biggl(R'_G
         +R^B_{GG} 
         +R^A_{GG} 
         -\int\limits^1_{x} dz
                               \bar \Peu_{GG}^{\Theta A} (z)
          \biggr)\Biggr].
\end{split}
\end{equation}
Keeping in mind that $R^B_{qq}=0$, we recover in eq.~(\ref{eq:sigmabar}) 
the complete {\em virtual} form factor $R_k$ 
\begin{equation}
  R'_k+R^A_{kk}+R^B_{kk}=R_k=(\tau-\tau_0)^{-1}\Phi_k(\tau,\tau_0) 
   =\frac{\alpha_S(t_A)}{\pi} \bigg[
     B_{kk} \ln\frac{1}{\epsilon}
    -A_{kk}    \bigg],
\end{equation}
see eq.~(\ref{eq:Rkkexpansion}).
Finally we arrive at the following expression:
\begin{equation}
\label{eq:sigmabar2}
  \begin{split}
  \bar\sigma_q
  &=\int\limits_0^1 dx 
    \int\limits^1_x dZ\;
  H_q(x)\; 
   A_{q} x^{-1+\eta}\;
      e^{-(\tau-\tau_0)
          \bigl(R_q 
         -\bar\Omega^A_{q} (x)
          \bigr)}
\\ &
   = \bar\sigma^{(a)}_q,\;\;\; \bar\sigma^{(b)}_q=0,
\\
  \bar\sigma_G
  &=\int\limits_0^1 dx 
    \int\limits^1_x \frac{dZ}{Z}\;
  H_G(x)\; 
   A_{G} x^{-1+\eta}\;
      e^{-(\tau-\tau_0)
        \bigl(R_G
         -\bar\Omega^A_{G} (x/Z)
          \bigr)}
       \hat d^B_G(\tau, Z|\tau_{0})
\\ &
   +\int\limits_0^1 dx\;   H_G(x)\; 
   A_{G} x^{-1+\eta}\;
      e^{-(\tau-\tau_0)
       \bigl(R_G
         -\bar\Omega^A_{G} (x)
          \bigr)}
\\ &
   = \bar\sigma^{(a)}_G +\bar\sigma^{(b)}_G.
\end{split}
\end{equation}

\begin{figure}[!ht]
  \centering
  {\epsfig{file=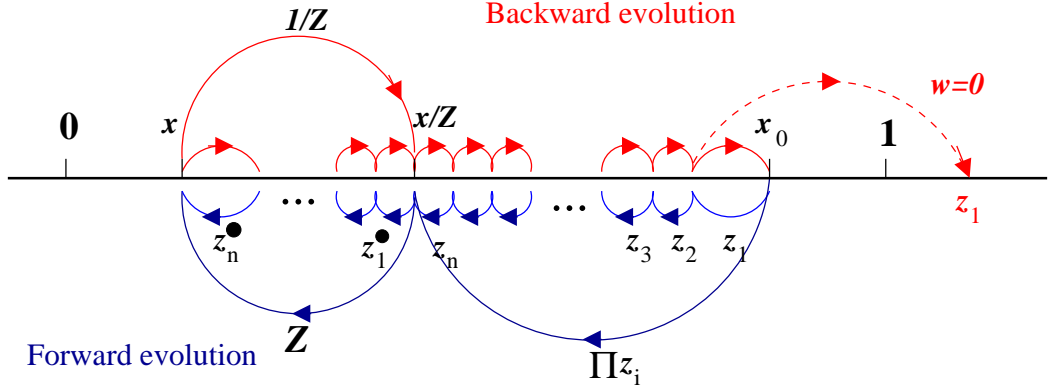,width=140mm}}
  \caption{\sf
    Scheme of the II.b MC algorithm in x space
    before relabelling, for the case of pure bremsstrahlung.
    $x_0=x/(Z\prod_i z'_i)$ results from $Z$ and $\prod z_i$.
    The zero-weight case of $x_0>1$ is also indicated.
    }
  \label{fig:FBscheme}
\end{figure}

The MC algorithm of type II.b for generating 
single (weighted) MC event consists of the following steps:
\begin{enumerate}
\item
  Generate a {\em branch index} $X=a,b$ according to a probability
  proportional to $\bar\sigma_k^{(X)}$; {\tt FOAM} does that efficiently.
\item
  For given $X$ generate variables $x$ and $Z$ or only $x$
  according to the integrand of the corresponding integral $\bar\sigma_k^{(X)}$;
  also done by {\tt FOAM}.
\item
  In the case $X=a$ generate two emission multiplicities $n'$ and $n^{\bu}$,
  the first one according to the Poisson distribution with 
  $\langle n'\rangle =(\tau-\tau_0)\Omega^A_k(x/Z)$
  and the other one according to the Bessel-type distribution with 
  $\lambda=(\tau-\tau_0)R^B_{GG}\ln(1/Z)$
  (as in the toy models).
\item
  Knowing the multiplicities, generate the variables
  $(\tau'_i,z'_i),~ i=1,\dots,n'$ and 
  $(\tau^{\bu}_j,z^{\bu}_j),~ j=1,\dots,n^\bu$,
  using methods described earlier.
\item
  Relabel the emission vertices, guided by the order of the $\tau$ variables.
\item
  Calculate the final MC weight, the same as was neglected at the early stage
  of generating ``phase-space'' variables.
\end{enumerate}
The above algorithm 
is also illustrated schematically in  fig.~\ref{fig:FBscheme}, 
in the $x$-space, before the {\em relabelling}.
Arrows help to understand the order of the reconstruction of all
$x$ variables out of $z$ variables.

\begin{figure}[!ht]
  \centering
  {\epsfig{file=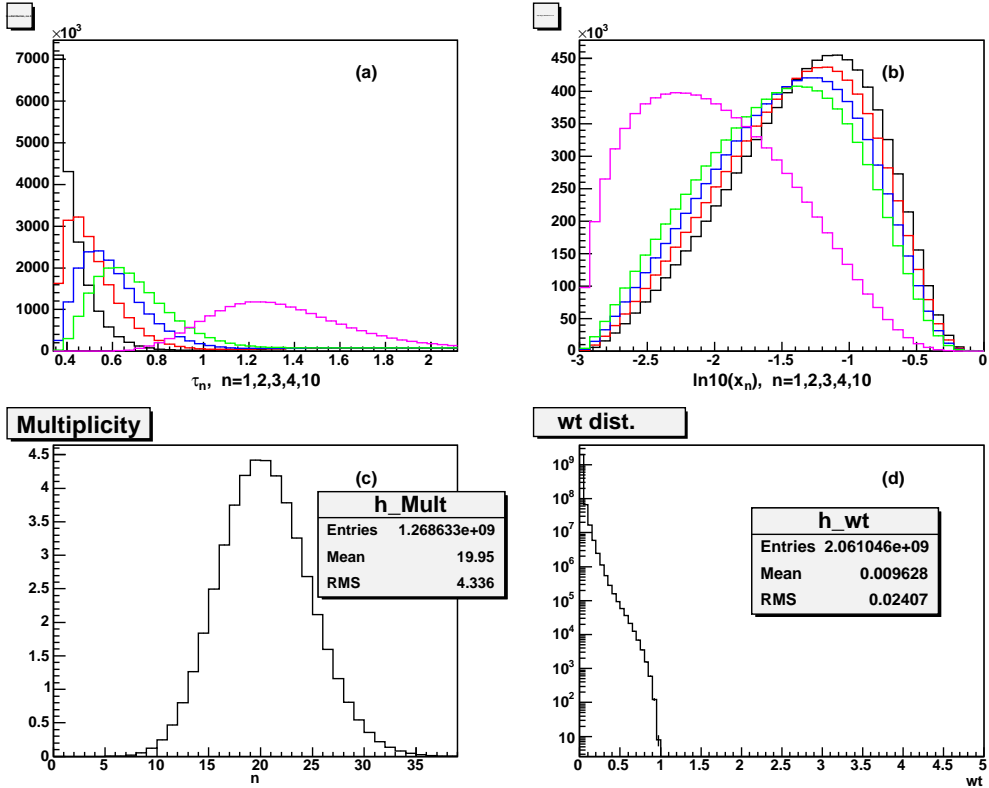,width=140mm}}
  \caption{\sf
    Results form {\tt GenIIb} MC prototype for gluon emitter, $k=G$.
    }
  \label{fig:IIB1gg}
\end{figure}
\begin{figure}[!ht]
  \centering
  {\epsfig{file=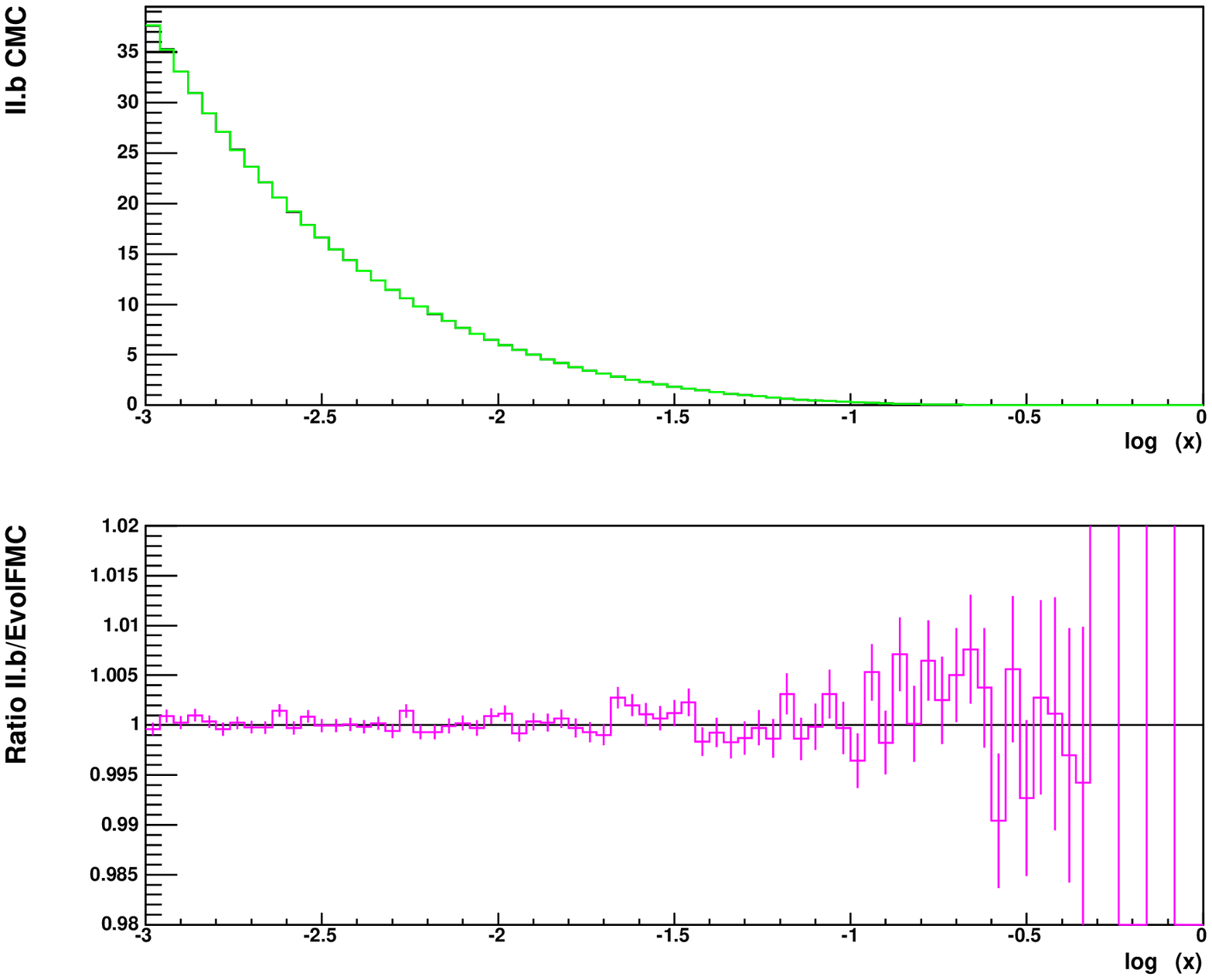,width=140mm}}
  \caption{\sf
    Comparison of II.b MC {\tt GenIIb} with the Markovian MC {\tt EvolMC}
    for pure gluonstrahlung $k=G$, $D_G(\tau_0,z)$ as in proton. $N_f=0$.
    }
  \label{fig:IIB-Mark-compar}
\end{figure}

In fig.~\ref{fig:IIB1gg} we show numerical results for the II.b prototype MC
for the same semi-exclusive $x$- and $\tau$-distributions as previously.
MC results coincide very well with
these from {\tt BremsP} in figs.~\ref{fig:BremsPggProt}.
This is a highly non-trivial result, having in mind sophistication
of the algorithm II.b.
Let us stress that the above agreement cannot be obtained without
a correct {\em relabelling} procedure being performed in the final stage of 
the algorithm II.b%
\footnote{ We have checked this fact numerically in a separate MC exercise.}
.

The generation time of an event (before any rejections)
is similar for both algorithms 
II.a and II.b. Therefore the acceptance, i.e.\ the ratio of the average 
to maximum weight, is a good measure of the overall efficiency of the 
algorithms.
The acceptance
for the new algorithm type II.b, as read from the weight distribution
in fig.~\ref{fig:IIB1gg} is $9.6\times 10^{-3}$.
This is a little bit worse than expected; it is, however, fully satisfactory 
-- it is better by a factor of $500$ than  the efficiency $2\times 10^{-5}$ 
for the solution II.a (without multibranching),
see fig.~\ref{fig:BremsPggProt}.
Using algorithm II.a as a guide, one may argue that
the $9.6\times 10^{-3}$ efficiency can still be improved
by a factor of 2 by including exactly the $x_0^{-0.8}$ factor.
Another factor of 3 could be obtained by performing a modelling 
of the $(1-x_0)^5$ distribution in an internal rejection loop of the algorithm.
In this way the overall efficiency
may go up to the level of 5\%.

Plots in fig.~\ref{fig:IIB1gg} show tests of exclusive distributions and 
efficiency, but not the overall normalization.
A strong test of the overall normalization of the algorithm II.b
is shown in fig.~\ref{fig:IIB-Mark-compar}, where high statistics
($\sim 4\times 10^9$ events) results of the II.b MC are compared with 
those of the
forward Markovian MC {\tt EvolMC} of ref.~\cite{Jadach:2003bu}%
\footnote{ Results of {\tt EvolMC} were in turn cross-checked 
  very precisely with the results
  of two non-MC evolution programs; see ref.~\cite{Jadach:2003bu}.}.
The agreement is reached within a statistical error of about 0.1\%
for $x<0.01$ and of 0.3\% for $x<0.1$. For higher $x$, in spite of
the extreme smallness of $D_G(x)$ (over 9 orders of magnitude),
the agreement holds perfectly well within the statistical errors.

{\em At this point we may state that our method II.b 
of solving the constrained MC really works in practice
and is reasonably efficient.}

We want to stress that it would not be possible to reach this conclusion 
without constructing and testing the explicit prototype of the algorithm 
type II.a, and other auxiliary MC exercises, as we did in this work.

In the above numerical exercises we have restricted ourselves to the LL case,
with $N_f=3$ massless quarks.
The QCD evolution kernels are unique and well known, and we therefore skip their
explicit definition.
The running constant
$\alpha_S(t)= 2\pi/( \beta_0 (t-\ln\Lambda_0))$
was used with $\Lambda_0= 0.245748338$.
The following starting values of the parton distributions in proton 
at $Q_0=1$ GeV were used in all our numerical exercises:
\begin{equation}
  \begin{split}
    xD_G(Q_0,x)        &= 1.9083594473\cdot x^{-0.2}(1-x)^{5.0},\\
    xD_q(Q_0,x)        &= 0.5\cdot xD_{\rm sea}(x) +xD_{2u}(x),\\
    xD_{\bar q}(Q_0,x) &= 0.5\cdot xD_{\rm sea}(x) +xD_{d}(x),\\
    xD_{\rm sea}(Q_0,x)&= 0.6733449216\cdot x^{-0.2}(1-x)^{7.0},\\
    xD_{2u}(Q_0,x)     &= 2.1875000000\cdot x^{ 0.5}(1-x)^{3.0},\\    
    xD_{d}(Q_0,x)      &= 1.2304687500\cdot x^{ 0.5}(1-x)^{4.0}.   
  \end{split}
\end{equation}

\section{Summary}
In this paper we presented a MC algorithm, which belongs to a new class of MC
algorithms  capable of generating a constrained Markovian evolution of 
parton distributions
according to DGLAP evolution equations.
Practical numerical implementation is for the moment restricted to the pure
bremsstrahlung case. 
Since the algebraic framework is defined for the full DGLAP,
it is therefore a matter of more programming
to extend it to the general case.
In the presented numerical tests (pure bremsstrahlung)
the algorithm has been checked against the dedicated forward evolution
(Markovian) MC program that we have written. 
We found an agreement at the level of 0.1\%.
The measured efficiency of the constrained MC is found to be quite satisfactory.
This work opens the way to a new class of MC algorithms
with possible applications in the initial-state QCD parton shower MC.
Furthermore, the Bessel-type distribution of the number of emissions, 
which forms the core of our algorithms, 
is similar to this obtained from the CCFM approach 
\cite{CCFM}, as shown in 
\cite{Kharraziha:1997dn}. This suggests another possible area of applications
of our algorithms.

\vspace{10mm}
\noindent
{\bf Acknowledgements}\\
We would like to thank W. P\l{}aczek and T. Sj\"ostrand for useful 
discussions.
We thank, for their warm hospitality, the CERN Particle Theory Group, 
where part of this work was done.

\appendix

\vfill\newpage

\noindent
{\bf\large APPENDIX}

\section*{The technique of the kernel split (multibranching)}

In section \ref{sect:II.b} we have shown how to reorganize integration
variables in the evolution iterative solution, such that in
the Monte Carlo integration/simulation algorithm it is possible
to generate first the chain of flavour indexes (gluon or quark type)
and the corresponding evolution time variables $\tau_i$
(i.e. those of the emissions which change flavour),
and later the other variables corresponding to gluon
emissions (no flavour change).

\begin{figure}[!h]
  \centering
  \epsfig{file=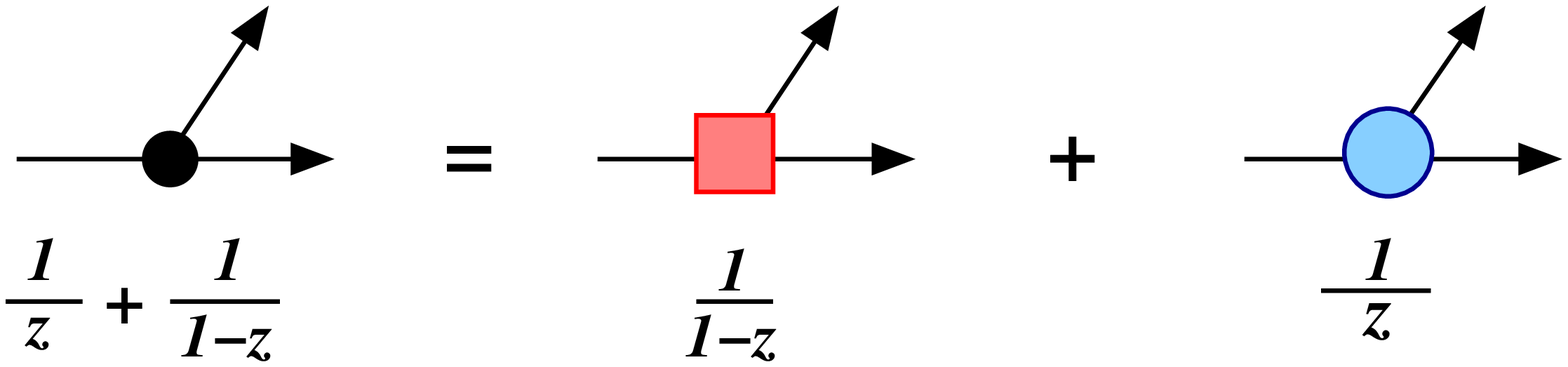,width=80mm}
  \caption{\sf
    Graphical representation of the split of the (approximate) 
    gluon emission kernel into two parts.
    }
  \label{fig:split}
\end{figure}
In the following we shall describe the application of the
 MC technique of multibranching to our problem.
In the multibranching one splits the integrand into many positive
components, chooses randomly one at a time and generates points
according to this particular component.
In the context of the iterative solution of the evolution equation,
it is worthwhile to apply this technique to the kernel for the transition
of the gluon into gluon:
\begin{equation}
  P_{GG}(z)=2C_A \left[ 
              \frac{z}{1-z} +\frac{1-z}{z} +z(1-z) \right],
\end{equation}
which has two very different singularities $1/z$ and $1/(1-z)$.
Therefore it is profitable in the Monte Carlo to
split $P_{GG}(z)=P^A_{GG}(z)+P^B_{GG}(z)$ such that 
$P^{A}_{GG}(z) \sim 1/(1-z)$ and $P^B_{GG} \sim 1/z$
(see fig.~\ref{fig:split}),
and to generate them separately, applying 
additional MC methods suited
to the individual character of each type of singularity%
\footnote{One should also take care of the positivity of the two components. 
  For simplicity we would like to have $P^B_{GG}(z)=2C_A/z$. 
  However, in such a case $P^A_{GG}(z)=P_{GG}(z)-P^B_{GG}(z)$ 
  is not positive.
  A possible solution is to first simplify
  $P_{GG}(z)\to \bar P_{GG}(z)= 2C_A [1/z + 1/(1-z)]$,
  compensating the simplification
  with the MC weight at a later stage, and then to split $\bar P_{GG}(z)$
  into two positive components without any problem.
  We shall come back to this point later on.}.

Since we already know from section \ref{sect:II.b} how to isolate
the pure brems\-strahlung subintegrals $d'_k$; see eq.\
(\ref{eq:dfun}), let us concentrate on one of them:
\begin{equation}
  \label{eq:d-funct} 
  \begin{split}
  d'_k(\tau,x | \tau_0) 
  =& e^{-(\tau-\tau_0)R_{kk}} \delta(x-1)
  +\sum_{n=1}^\infty
   \Biggl(
   \prod_{j=1}^n \int\limits_{\tau_0}^\tau d\tau_j\; 
                 \Theta(\tau_j-\tau_{j-1})
                 \int\limits_0^1 dz_j
   \Biggr)
\\&\times
      e^{-(\tau-\tau_n)R_{kk}}
   \Biggl(
      \prod_{i=1}^n \Pcal_{kk}^\Theta (z_i)\;
                    e^{-(\tau_i-\tau_{i-1})R_{kk}}
   \Biggr)
      \delta\bigg(x- \prod_{i=1}^n z_i \bigg),
  \end{split}
\end{equation}
where in reality we are
interested in the gluon case $\Pcal_{kk}(z)=\Pcal_{GG}(z)$.

Taking advantage of the independence of the kernels on $\tau$
we can rewrite the above equation as follows:
\begin{equation}
  \label{eq:d-funct2}
  \begin{split}
  d'_k(\tau,x | \tau_0) 
  &= e^{-(\tau-\tau_0)R_{kk}}
   \sum_{n=0}^\infty \frac{(\tau -\tau_0)^n}{n!}
   \prod_{i=1}^n \int\limits_0^1 dz_i\;
\\&\times
      \prod_{i=1}^n 
         [\Pcal_{kk}^{\Theta A} (z_i) + \Pcal_{kk}^{\Theta B} (z_i)]\;
      \delta\bigg(x- \prod_{i=1}^n z_i \bigg),
  \end{split}
\end{equation}
where more compact notation is achieved
by defining $\prod\limits_{i=1}^0 \equiv 1$.
Note that at this stage we made certain important short-cuts, because
we have integrated over $\tau$. This simplifies the argument
but makes it questionable in view of certain important claims
concerning the final distribution in the space of 
$(n;\tau_1,z_1,\tau_2,z_2,...,\tau_n,z_n)$,
which we are going to make at the end of this appendix. 
We shall therefore refine our proof later on, showing
how to proceed for the distributions with unintegrated $\tau$'s.
\begin{figure}[!h]
  \centering
  \epsfig{file=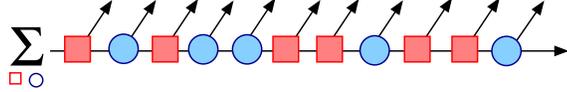,width=80mm}
  \caption{\sf
    Reorganization of the multiple gluon emission leading to multibranching.
    }
  \label{fig:mbran}
\end{figure}

Let us now reorganize the overall sum as follows %
(see also schematic illustration in fig.~\ref{fig:mbran}),
\begin{equation}
  \label{eq:d-funct3}
  \begin{split}
  d'_k(\tau,x | \tau_0) 
 &= e^{-(\tau-\tau_0)R_{kk}} 
   \sum_{n_1=0}^\infty \sum_{n_2=0}^\infty 
   \frac{(\tau -\tau_0)^{n_1+n_2}}{n_1!n_2!}
   \prod_{i=1}^{n_1} \int\limits_0^1 dz_i\;
   \prod_{j=1}^{n_2} \int\limits_0^1 dz_j\;
\\&\times
      \prod_{i=1}^{n_1} \Pcal_{kk}^{\Theta A} (z_i)\;
      \prod_{j=1}^{n_2} \Pcal_{kk}^{\Theta B} (z_j)\;\;
      \delta\bigg(x- \prod_{i=1}^{n_1} z_i \prod_{j=1}^{n_2} z_j \bigg),
  \end{split}
\end{equation}
where the two sums take care of the two kernel components.
We can now factorize the whole integral as a convolution
of the two integrals, each of them corresponding to one component of
the kernel:
\begin{equation}
\label{eq:dwa}
\begin{split}
&d'_k(\tau,x | \tau_0) =
\int_0^1 dz^A \int_0^1 dz^B \;
  \delta(x-z^A z^B)\;
   d'^A_k(\tau,z^A | \tau_0)\;
   d'^B_k(\tau,z^B | \tau_0),\\
& d'^X_k(\tau,x | \tau_0)=
  e^{-(\tau-\tau_0)R^X_{kk}}
  \sum_{n=0}^\infty
     \frac{(\tau -\tau_0)^{n}}{n!}
     \prod_{i=1}^{n} \Pcal_{kk}^{\Theta X} (z_i)\;
     \delta\bigg(x- \prod_{j=1}^{n} z_j \bigg),\;\;  X=A,B.
\end{split}
\end{equation}
The functions $R^X$ are constrained only by 
\begin{equation}
\sum_X R^X_{kk} = R_{kk}.
\end{equation}
For example in some cases they may be defined as
\begin{equation}
R^X_{kk} = \int_0^{1-\veps} dz z \Pcal_{kk}^{\Theta X} (z).
\end{equation}

We may restore the ordered evolution time integrals
\begin{equation}
d'^X_k(\tau,x | \tau_0)=
  e^{-(\tau-\tau_0)R^X_{kk}}
  \sum_{n=0}^\infty
  \prod_{j=1}^n \int\limits_{\tau_0}^\tau d\tau_j\; \Theta(\tau_j-\tau_{j-1})
  \prod_{i=1}^{n} \Pcal_{kk}^{\Theta X} (z_i)\;
  \delta\bigg(x- \prod_{j=1}^{n} z_j \bigg).
\end{equation}
\begin{figure}[!h]
  \centering
  \epsfig{file=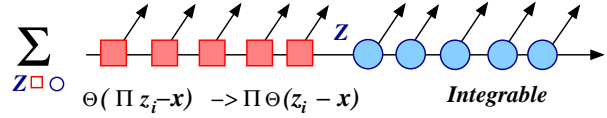,width=80mm}
  \caption{\sf
    Further reorganization of the multiple gluon emission in the multibranching.
    }
  \label{fig:mbran2}
\end{figure}
However, it should be remembered that the variables $\tau_i$ and $z_i$
are not exactly the same as in the original integral (in spite of the same
notation) but they are related 
by means of a ``relabelling'' procedure described later in this appendix.
The above algebra is represented schematically in fig.~\ref{fig:mbran2}.

It is now possible to implement the integral of eq.~(\ref{eq:dwa})
as a pair of two independent ``parallel Markovian processes'',
both starting at $\tau_0$ and stopping at $\tau$.
The first one would have decay constant $R^A_{kk}$
and variable $z_i$ generated according to $\Pcal_{kk}^{\Theta A} (z_i)$, 
yielding emission multiplicity $n_1$ at the stopping point,
while the second one would have its
decay constant $R^B_{kk}$, variables $z_j$ generated according to
$\Pcal_{kk}^{\Theta B} (z_j)$, and the emission multiplicity $n_2$.

\begin{figure}[!h]
  \centering
  \epsfig{file=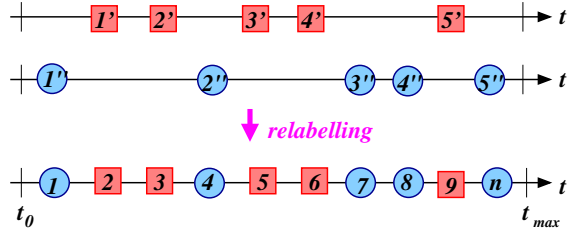,width=80mm}
  \caption{\sf
    Illustration of ``relabelling''.
    The actual generation is done in two steps:
    First, for each of the two branches (squares and circles) 
    the ordered $\tau$'s are generated separately and independently
    in the entire $\tau$-range.
    Next, $(\tau_i,z_i)$ are {\em relabelled} according 
    to a {\em common} ordering in $\tau$.
    Only after such a relabelling is $x$ constructed:
    $x=\prod^n_{j=1} z_j$.
    }
  \label{fig:mbran3}
\end{figure}
It is important to understand that at the very end 
the two sets of $z_i,\tau_i$, $i=1,...,n_1$
and $z_j,\tau_j$, $j=1,...,n_2$ can be {\em merged},
forgetting from which parallel generation branch they originate.
Merging is done simply by creating a common list of ordered
variables $\tau_i$ and renaming/reordering $z_i$ variables in
{\em exactly} the same way. 
Such {\em relabelling} procedure will undo
the procedure of combining together the 
$\left( {n_1+n_2 \atop n_1 } \right)$ terms
done in eq.~(\ref{eq:d-funct3}).
The relabelling procedure is illustrated schematically in fig.~\ref{fig:mbran3}.
The resulting $z_l$, $l=1,...,n_1+n_2$ will be then distributed
according to the product
\begin{equation}
\prod_{l=1}^{n_1+n_2}
 [\Pcal_{kk}^{\Theta A} (z_l)+\Pcal_{kk}^{\Theta B} (z_l)]
\end{equation}
Moreover, also the total multiplicity $n=n_1+n_2$ and 
the evolution times $\tau_l$, $l=1,...,n$ will
be distributed as if they were coming from the
corresponding single Markovian MC.

Actually, the reader may be concerned that the above claim is not really
founded on a solid derivation because we have excluded the $\tau$
space in the binomial decomposition after integrating over $\tau$'s
at an early stage of derivation, while we are now making statements on
the distribution in the full space $\tau_1,z_1,\tau_2,z_2,...,\tau_n,z_n$.
We need clearly to refine our derivation keeping the $\tau$-space alive.
The full derivation involves non-trivial combinatorics,
and here we shall only give a sketch on the necessary reasoning.
Consider the expression with three kernels
\begin{equation}
\begin{split}
&\int d\tau_1 d\tau_2 d\tau_3 \Theta_{321}
\int dz_1 dz_2 dz_3
 (A(3)+B(3))(A(2)+B(2))(A(1)+B(1)),
\end{split}
\end{equation}
where we abbreviate:
$\Theta_{321}=\Theta(\tau_3-\tau_2)\Theta(\tau_2-\tau_1)$
and $A(i)=\Peu_{kk}^A(z_i)$, $B(j)=\Peu_{kk}^B(z_j)$.
It is decomposed as follows 
\begin{equation}
\begin{split}
\int d\tau_1 d\tau_2 d\tau_3 \Theta_{321} \int dz_1 dz_2 dz_3
  & [A(3)A(2)A(1)
\\& +B(3)A(2)A(1)+A(3)B(2)A(1)+A(3)A(2)B(1)
\\& +B(3)B(2)A(1)+B(3)A(2)B(1)+A(3)B(2)B(1)
\\& +B(3)B(2)B(1)]
\end{split}
\end{equation}
Each of the four groups in four lines is now transformed
separately into a single factor with different ordering pattern
of the $\tau$ variables. For instance the second line we
transform explicitly as follows:
\begin{equation}
\begin{split}
\int& d\tau_1 d\tau_2 d\tau_3 \Theta_{321} \int dz_1 dz_2 dz_3
     [B(3)A(2)A(1)+A(3)B(2)A(1)+A(3)A(2)B(1)]
\\&=\int d\tau_1 d\tau_2 d\tau_{1'} dz_1dz_2dz_{1'} \Theta_{1'21} B(1')A(2)A(1)
\\&+\int d\tau_1 d\tau_{1'} d\tau_3 dz_1dz_{1'}dz_3 \Theta_{31'1} A(3)B(1')A(1)
\\&+\int d\tau_{1'} d\tau_2 d\tau_3 dz_{1'}dz_2dz_3 \Theta_{321'} A(3)A(2)B(1')]
\end{split}
\end{equation}
where we essentially renamed both $z$'s and $\tau$'s sitting in the $B$-factor.
The same can be done for variables in the $A$-factors: 
\begin{equation}
\begin{split}
   &\int d\tau^\bu_1 d\tau^\bu_2 d\tau'_1      dz^\bu_1 dz^\bu_2 dz'_1 
                            \Theta_{1' 2^\bu 1^\bu} B(1')A(2^\bu)A(1^\bu)
\\+&\int d\tau^\bu_1 d\tau'_1 d\tau^\bu_2  dz^\bu_1 dz'_1 dz^\bu_2 
                            \Theta_{2^\bu 1' 1^\bu} A(2^\bu)B(1')A(1^\bu)
\\+&\int d\tau'_1 d\tau^\bu_1 d\tau^\bu_2      dz'_1 dz^\bu_1 dz^\bu_2 
                            \Theta_{2^\bu 1^\bu 1'} A(2^\bu)A(1^\bu)B(1')].
\end{split}
\end{equation}
Let us summarize 
explicitly the {\em relabelling} of the variables that has been done above:
\begin{equation}
\begin{split}
\hbox{\rm for}~~~ \tau'_1         > \tau^\bu_2         > \tau^\bu_1:~~~~
         & \tau_1=\tau'_1,\; \tau_2=\tau^\bu_2,\; \tau_1=\tau^\bu_1,~~
              z_1=   z'_1,\;    z_2=   z^\bu_2,\;    z_1=   z^\bu_1,\\
\hbox{\rm for}~~~ \tau^\bu_2         > \tau'_1         > \tau^\bu_1:~~~~
         & \tau_1=\tau^\bu_2,\; \tau_2=\tau'_1,\; \tau_1=\tau^\bu_1,~~
              z_1=   z^\bu_2,\;    z_2=   z'_1,\;    z_1=   z^\bu_1,\\
\hbox{\rm for}~~~ \tau^\bu_2         > \tau^\bu_1         > \tau'_1:~~~~
         & \tau_1=\tau^\bu_2,\; \tau_2=\tau^\bu_1,\; \tau_1=\tau'_1,~~
              z_1=   z^\bu_2,\;    z_2=   z^\bu_1,\;    z_1=   z'_1,
\end{split}
\end{equation}
Now, we may pull out the kernels and combine the $\Theta$-functions
\begin{equation}
\begin{split}
   \int &d\tau^\bu_1 d\tau^\bu_2 d\tau'_1 dz^\bu_1dz^\bu_2dz'_1
    B(1')A(2^\bu)A(1^\bu)[\Theta_{1'2^\bu 1^\bu}
                        + \Theta_{2^\bu 1'1^\bu} +\Theta_{2^\bu 1^\bu 1'}]
\\& =
   \int d\tau^\bu_1 d\tau^\bu_2 d\tau'_1\; dz^\bu_1dz^\bu_2 dz'_1\;\;
    \Theta_{2^\bu 1^\bu} A(2^\bu)A(1^\bu) \;\; \Theta'_1 B(1'),
\end{split}
\end{equation}
where $\Theta_{1'}=1$ and $\Theta_{2^\bu 1^\bu}=\Theta(\tau^\bu_2-\tau^\bu_1)$.
The above tedious relabelling of $z$'s and $\tau$'s and recombining
of $\Theta$'s into product of two independent ones can be done
for any number of kernels. The net result is an interesting identity:

\begin{equation}
\label{eq:identity}
\begin{split}
 \prod_{i=1}^n \int_{\tau_{i-1}}^1 d\tau_i
 &\int_0^1 dz_i\;
  (A(z_i)+B(z_i)) w({\bf \tau}, {\bf z})
\\=\sum_{n_1=0}^n
 &\left(
  \prod_{i=1}^{n_1} \int_{\tau^\bu_{i-1}}^1 d\tau^\bu_i 
  \int_0^1 dz^\bu_i\;
  A(z^\bu_i)
  \right)\;\;
  \left(
  \prod_{i=1}^{n-n_1} \int_{\tau'_{i-1}}^1 d\tau'_{i}\;
  \int_0^1 dz'_{i}\;
  A(z'_{i}) \right)
\\
  &w\bigl({\bf \boldtau(\boldtau^\bu,\boldtau')}, 
    {\bf z(\boldtau^\bu;\boldtau',z^\bu,z')}\bigr), 
\end{split}
\end{equation}
where $w(\boldtau, {\bf z})$ is an arbitrary ``test function''
ensuring that eq.\  (\ref{eq:identity}) is indeed a {\em differential}
identity, and not an obvious multiplication rule of exponential functions. 
The mapping (relabelling) $\tau_i=\tau_i({\bf \boldtau^\bu,\boldtau'})$ and 
$z_i=z_i({\bf \boldtau^\bu;\boldtau',z^\bu,z'})$ is nothing
more than a permutation of the integration variables,
which is ``guided'' by the ordering of the $\tau$ variables,
much as in the explicit example above.
Note that the above identity is still valid if the integrand involves
any additional factor, symmetric with respect to the permutation
of the integrand variables $\tau_i$ and $z_i$.

The above formula is a kind of generalization of the Newton
binomial formula in which an $n$-dimension simplex in $\tau$ variables
is decomposed into a sum over the Cartesian product of the
two simplexes in $n_1$ and $n-n_1$ dimensions.
From this exercise it is also clear that this identity
implicitly involves a relabelling of $z$ variables depending
on the ordering of $\tau$ variables. This is exactly
what we have to do in the Monte Carlo if we generate $A(z)$
and $B(z)$ independently, but we want to have the distribution
$A(z)+B(z)$ at the end of the algorithm.
Note that a similar MC procedure with relabelling of the integration variables
was done in the context of the ISR and FSR photon radiation in the YFS3
algorithm of KKMC, before adding the ISR--FSR interference \cite{Jadach:2000ir}.


\providecommand{\href}[2]{#2}

\end{document}